\documentclass[sigconf]{acmart}
\usepackage{graphicx}
\usepackage{multirow}
\usepackage{makecell}
\usepackage{listings}
\usepackage{framed}
\usepackage{hyperref}
\usepackage[capitalise,noabbrev]{cleveref}
\usepackage{dblfloatfix}
\usepackage{placeins}
\usepackage{soul}
\usepackage{microtype}
\usepackage{todonotes}
\usepackage{xspace}
\usepackage{subcaption}

\AtBeginDocument{%
  }

\newcolumntype{L}[1]{>{\raggedright\let\newline\\\arraybackslash\hspace{0pt}}m{#1}}
\newcolumntype{C}[1]{>{\centering\let\newline\\\arraybackslash\hspace{0pt}}m{#1}}
\newcolumntype{R}[1]{>{\raggedleft\let\newline\\\arraybackslash\hspace{0pt}}m{#1}}

\newcommand{\code}[1]{\texttt{\small #1}}
\renewcommand{\paragraph}[1]{\vspace{6pt}\noindent{\bf #1}\hspace{6pt}}

\usepackage{tikz}
\usetikzlibrary{arrows}
\usetikzlibrary{arrows.meta}
\usetikzlibrary{patterns}
\usepackage{pgfplots}
\usepgfplotslibrary{groupplots}
\usepgfplotslibrary{statistics}

\definecolor{color-a}{RGB}{244, 241, 222}
\definecolor{color-b}{RGB}{129, 178, 154}
\definecolor{color-c}{RGB}{61, 64, 91}
\definecolor{color-d}{RGB}{242, 204, 143}
\definecolor{color-e}{RGB}{224, 122, 95}
\definecolor{color-f}{RGB}{201, 228, 202}
\definecolor{color-g}{RGB}{254, 217, 183}

\definecolor{gray}{gray}{0.9}
\definecolor{light-gray}{gray}{0.4}

\newcommand{\tool}{\textsc{GenSIaC}\xspace}

\newcommand{\codegen}{CodeGen2.5}
\newcommand{\starcoder}{StarCoder}
\newcommand{\wizardcoder}{WizardCoder}
\newcommand{\codellama}{CodeLlama}
\newcommand{\vicuna}{Vicuna}
\newcommand{\llama}{Llama2}
\newcommand{\gpt}{GPT3.5}
\newcommand{\gptfour}{GPT4}

\newcommand{\admdef}{\emph{Admin by Default}}
\newcommand{\emppwd}{\emph{Empty Password}}
\newcommand{\hardcd}{\emph{Hard-Coded Secret}}
\newcommand{\invdip}{\emph{Unrestricted IP Address}}
\newcommand{\suscmt}{\emph{Suspicious Comments}}
\newcommand{\notls}{\emph{Use of HTTP Without TLS}}
\newcommand{\weakalg}{\emph{Use of Weak Cryptography Algorithms}}
\newcommand{\noitchk}{\emph{No Integrity Check}}
\newcommand{\nodefsw}{\emph{Missing Default in Case Statement}}

\newcommand{\mycolorbox}[2]{{\sethlcolor{#1}\hl{\ \ \ \ \ \ }\;\!}}
\newcommand{\basebox}{\mycolorbox{color-e}}
\newcommand{\partialbox}{\mycolorbox{color-f}}
\newcommand{\gensiacbox}{\mycolorbox{color-b}}
\newcommand{\gptbox}{\mycolorbox{gray}}

\newcommand\realnumberstyle[1]{}
\makeatletter
\newcommand{\linecolor}[3]{
    {\realnumberstyle{#3}}
    \begingroup
    \lst@basicstyle
    \ifnum\value{lstnumber}=#1
        \color{#2}
    \else
        \color{white}
    \fi
    \rlap{\hspace*{\lst@numbersep}
    \color@block{\linewidth}{\ht\strutbox}{\dp\strutbox}
    }
    \endgroup
}
\makeatother

\setcopyright{acmlicensed}
\copyrightyear{2024}
\acmYear{2024}
\acmDOI{XXXXXXX.XXXXXXX}

\author{Yikun Li$^{*}$, Matteo Grella$^{*}$, Daniel Nahmias$^{\dagger}$, Gal Engelberg$^{\dagger}$, Dan Klein$^{\dagger}$, Giancarlo Guizzardi$^{*}$, Thijs van Ede$^{*}$, Andrea Continella$^{*}$}
\affiliation{%
  \institution{$^{*}$University of Twente}
  \city{Enschede}
  \country{Netherlands}
}
\affiliation{%
  \institution{$^{\dagger}$Accenture Labs}
  \city{Tel Aviv}
  \country{Israel}
}

\renewcommand{\shortauthors}{Li et al.}

\begin{document}

\title{\tool{}: Toward Security-Aware Infrastructure-as-Code Generation with Large Language Models}

\begin{abstract}
In recent years, \emph{Infrastructure as Code (IaC)} has emerged as a critical approach for managing and provisioning IT infrastructure through code and automation. IaC enables organizations to create scalable and consistent environments, effectively managing servers and development settings. However, the growing complexity of cloud infrastructures has led to an increased risk of misconfigurations and security vulnerabilities in IaC scripts. To address this problem, this paper investigates the potential of \emph{Large Language Models (LLMs)} in generating security-aware IaC code, avoiding misconfigurations introduced by developers and administrators.

While LLMs have made significant progress in natural language processing and code generation, their ability to generate secure IaC scripts remains unclear. This paper addresses two major problems: 1) the lack of understanding of security weaknesses in IaC scripts generated by LLMs, and 2) the absence of techniques for enhancing security in generating IaC code with LLMs.

To assess the extent to which LLMs contain security knowledge, we first conduct a comprehensive evaluation of base LLMs in recognizing major IaC security weaknesses during the generation and inspection of IaC code. 
Then, we propose \tool{}, an instruction fine-tuning dataset designed to improve LLMs' ability to recognize potential security weaknesses. 
Leveraging \tool, we fine-tune LLMs and instruct models to generate security-aware IaC code. 
Our evaluation demonstrates that our models achieve substantially improved performance in recognizing and preventing IaC security misconfigurations, e.g., boosting the F1-score from 0.303 to 0.858.
Additionally, we perform ablation studies and explore \tool{}'s generalizability to other LLMs and its cross-language capabilities.
\end{abstract}

\maketitle

\section{Introduction}

Infrastructure as Code (IaC) is an approach to managing and provisioning IT infrastructure through the use of code and automation, rather than manual processes and configuration.
In recent years, IaC has gained significant traction in the realm of DevOps due to its ability to create scalable and consistent environments, effectively managing servers and development settings \cite{morris2016infrastructure, rahman2019seven}.
Nowadays, IaC scripts play a crucial role in effectively managing servers and development settings.
For instance, Rahman \textit{et al.} \cite{rahman2019seven} reveal that Intercontinental Exchange (ICE), a prominent Fortune 500 organization, utilizes IaC scripts to manage 75\% of its 20,000 servers.
By employing IaC scripts, ICE has successfully reduced the time required to set up development environments from 1-2 days to a mere 21 minutes.

However, the growing complexity of cloud infrastructures has led to an increased risk of misconfigurations in IaC scripts, which can harbor flaws that lead to severe security consequences~\cite{dns_outage, rahman2019seven}.
For example, GitHub faced a DNS infrastructure outage due to IaC script errors \cite{dns_outage}. 
Besides, just like any type of code, IaC scripts can also be susceptible to security issues \cite{rahman2019seven}.
An example of IaC security weakness (i.e., \hardcd{}, CWE-798) is presented in \cref{fig:example}.
The code snippet shows an IaC script that creates a MySQL user. 
The security weakness arises from the use of hard-coded credentials, where the password is set to \emph{1234}.
This practice makes the system vulnerable to unauthorized access, while it is recommended to use a secure method for managing and storing sensitive information, such as environment variables or a secrets management system.
More in general, Saavedra and João \cite{DBLP:conf/kbse/Saavedra022} conducted a large-scale empirical study and found that 10.4\% to 40.1\% of IaC scripts contain at least one security weakness for different projects.
Furthermore, Rahman and Laurie \cite{DBLP:journals/ieeesp/RahmanW21} identified that 17.9\% to 32.9\% of the IaC scripts include at least one security weakness in IaC scripts developed using Ansible, Chef, and Puppet.
Even when a security weakness does not directly lead to a security breach, it requires inspection to prevent potential consequences.

To help developers deal with such complexity and mitigate misconfigurations, automating the IaC code generation process with security in mind has become a promising approach.
In recent years, Large Language Models (LLMs) have made remarkable progress in the field of natural language processing, as evidenced by groundbreaking works \cite{DBLP:conf/nips/VaswaniSPUJGKP17,brown2020language,DBLP:journals/corr/abs-2303-08774,touvron2023llama}. 
Concurrently, the training of LLMs on open-source code repositories has expanded their applicability, enabling LLMs to generate code based on user prompts with increasing accuracy and relevance \cite{DBLP:journals/corr/abs-2204-05999,li2023starcoder,nijkamp2023codegen2,DBLP:journals/corr/abs-2308-12950,DBLP:journals/corr/abs-2202-13169}.
As a result, these models have become the foundation for a growing number of code-generation tools available in the market, such as Copilot, CodeWhisperer, and GhostWriter \cite{copilot,codewhisperer,ghostwriter}.
Empirical research has consistently demonstrated the positive impact of LLMs on enhancing programming efficiency \cite{DBLP:conf/chi/Vaithilingam0G22,DBLP:conf/promise/YetistirenOT22}.

However, while LLMs demonstrate impressive functional correctness during code generation, they can also generate code with potential security risks \cite{DBLP:journals/corr/abs-2107-03374,pearce2022asleep,DBLP:journals/corr/abs-2304-09655}. 
A recent study \cite{pearce2022asleep} explored the security of code generated by Copilot in 89 different scenarios and found approximately 40\% of the generated code to be vulnerable.
Another work \cite{DBLP:journals/corr/abs-2304-09655} found that ChatGPT generated code that failed to meet minimum security standards in 16 out of 21 cases.

\begin{figure}[t]
\begin{lstlisting}[language=Java, numbers=left, numberstyle=\linecolor{4}{color-g}]
- name: Create a MySQL user.
  mysql_user:
    name: ""{{ domain }}""
    password: ""1234""
    priv: ""{{ domain }}.*:ALL""
    host: localhost
    state: present
\end{lstlisting}
\caption{Simplified example of \hardcd{} in an IaC script.}
\label{fig:example}
\vspace{-3mm}
\end{figure}

In this work, we address these challenges and create security-aware models for generating IaC code.
We focus on two significant problems: 1) there is no research investigating to what extent LLMs can recognize IaC security weaknesses while generating or inspecting IaC scripts; 2) there is no approach available to improve security awareness in LLMs during code generation and inspection. 
By developing security-aware models, we aim to provide developers with a more reliable and secure means of automating IaC script generation, ultimately reducing the risk of misconfigurations and security vulnerabilities in cloud infrastructures.
The two problems are detailed as follows.

\paragraph{Problem I: LLM-Generated IaC Code Lacks Security Awareness}
Prior research identified security flaws in IaC scripts \cite{rahman2019seven,DBLP:journals/ieeesp/RahmanW21,DBLP:conf/kbse/Saavedra022} that can lead to severe consequences \cite{dns_outage}. 
With the advancement of LLMs, generating IaC scripts has become increasingly popular \cite{AIaC,genIaC,genIaC2}.
However, the security of IaC code generated by LLMs has not been thoroughly investigated. 
For instance, if LLMs generate IaC scripts as shown in \cref{fig:example} without reminding the user to securely manage credentials, security weaknesses would arise in production systems.
However, the performance of LLMs in recognizing IaC security weaknesses remains unknown, which hinders their adoption during software development. 
In summary, there is an insufficient analysis of the security of IaC code generated by LLMs and the ability of LLMs to recognize security issues.

\paragraph{Problem II: Enhancing Security Awareness in LLMs for IaC Code Generation and Inspection}
Current state-of-the-art LLMs for code generation are primarily trained on code \cite{DBLP:journals/corr/abs-2204-05999,li2023starcoder,nijkamp2023codegen2,DBLP:journals/corr/abs-2308-12950,DBLP:journals/corr/abs-2202-13169}, with some also trained on a small portion of natural language documents. 
However, none of these models are specifically optimized to enhance security in generating IaC code. 
There are several key differences between generating IaC scripts and other types of code: 1) the security weaknesses are distinct; 2) due to the nature of IaC, many security weaknesses cannot be resolved without additional information from users, as illustrated in \cref{fig:example}.
This poses a severe problem when developers use the generated IaC code directly without understanding the potential security flaws present in the code. 
Additionally, it is problematic if developers rely on LLMs to identify security weaknesses in existing IaC scripts, but the models are not capable of doing so. 
Training LLMs is an extremely resource and time-intensive process, often requiring GPU clusters and extensive time investments ranging from weeks to months \cite{li2023starcoder,nijkamp2023codegen2,DBLP:journals/corr/abs-2308-12950}. 
Therefore, it is crucial to propose an approach that can improve the trained LLMs' ability to enhance security when generating or inspecting IaC code, using reasonable resources and time.

\paragraph{Our Solution}
To address the two aforementioned problems, we first investigate the performance of LLMs in recognizing security misconfigurations, also known as \emph{security weaknesses} or \emph{security smells}, during the generation and inspection of IaC code.
Second, we propose an approach to enhance the ability of existing LLMs to recognize such misconfigurations.

We begin by evaluating non-finetuned LLMs in recognizing nine major types of IaC security weaknesses using state-of-the-art frameworks \cite{opdebeeck2023control,saavedra2023polyglot}. Code-specific LLMs show poor F1-scores (0.000-0.303), while open-source general-purpose models (\vicuna{} and \llama{}) perform better (0.354-0.495). \gpt{} and \gptfour{} outperform them with F1-scores of 0.416-0.592.
Subsequently, we introduce \tool{}, a fine-tuning dataset to enhance LLMs' security weakness awareness in IaC scripts. After fine-tuning \codellama{} with \tool{}, F1-scores improve from 0.276-0.303 to 0.771-0.858, surpassing larger LLMs without fine-tuning (0.549-0.592, \gptfour{}). Syntactical and functional correctness scores also increase from 0.927 and 0.793 to 1.000 and 0.931, respectively.

\paragraph{Our Approach vs. Detection Techniques}
We utilize LLMs instead of traditional security weakness detection approaches (e.g., GLITCH~\cite{saavedra2023polyglot}) for recognizing security weaknesses during code generation due to the following reasons: 1) Ensuring that LLMs generate secure and reliable IaC scripts is crucial, as developers might inadvertently rely on error-prone code produced by these models; 2) Using GLITCH to detect security weaknesses after generating code with LLMs necessitates additional steps, making the process cumbersome for developers. In contrast, our method generates IaC code and reports vulnerabilities concurrently; 3) Our dataset allows all LLMs to be fine-tuned for generating more secure IaC code; 4) Developers who are not experts in IaC vulnerabilities may be unfamiliar with GLITCH, potentially hindering its adoption and effectiveness; 5) Our proposed approach offers greater extensibility, as it can generalize knowledge from learned IaC technologies and apply it to unknown IaC technologies, a capability that GLITCH currently lacks; 6) Our approach can combine the strengths of various tools (e.g., TOOL\_A detecting weakness A, TOOL\_B detecting weakness B) to create a comprehensive training dataset for fine-tuning and enhancing LLMs; 7) GLITCH relies on basic templates for vulnerability identification, which can lead to false predictions, such as mistaking 000000000 for an unrestricted IP address like 0.0.0.0. This highlights a deficiency in the general programming knowledge needed for precise detection.

\paragraph{Contributions} Overall, we make the following contributions:

\begin{itemize}
    \item A benchmark for evaluating the capability of models to recognize IaC security weaknesses during two tasks: IaC code generation and inspection.

    \item A comprehensive evaluation of base LLMs on various IaC security issues while generating and inspecting IaC code.

    \item \tool{}, a novel instruction tuning dataset, designed to enhance security awareness in LLMs for IaC code generation and inspection
    
    \item An extensive evaluation of models fine-tuned with \tool{} on a range of IaC security vulnerabilities, benchmarks, and LLMs.
\end{itemize}

In the spirit of open science, we make our data publicly available\footnote{\url{https://github.com/yikun-li/GenSIaC}}.

\section{Background and Related Work}

In this section, we present an overview of the background knowledge and prior research related to our study.

\subsection{Code Generation with LLMs}

LLMs have demonstrated significant potential in the area of code generation.
Numerous LLMs have been trained on extensive datasets, allowing them to generate code across various languages and use cases \cite{DBLP:journals/corr/abs-2204-05999,li2023starcoder,nijkamp2023codegen2,DBLP:journals/corr/abs-2308-12950,luo2023wizardcoder,vicuna2023,touvron2023llama,DBLP:conf/ccs/HeV23}. 
Among these models, some are primarily pre-trained on code repositories. 
Specifically, Fried \textit{et al.} \cite{DBLP:journals/corr/abs-2204-05999} introduced InCoder, a generative model of code capable of performing both program synthesis and editing through a causal masking objective. 
InCoder is trained to generate code from a vast corpus of permissively licensed code, encompassing 28 languages and 159 GB of code, as well as a corpus of 57 GB of StackOverflow questions, answers, and comments. 
Following this work, Li \textit{et al.} \cite{li2023starcoder} trained \starcoder{} using a large dataset of source code from The Stack v1.2, Jupyter notebooks, natural language conversations from GitHub issues, and Git commits. 
Subsequently, Nijkamp \textit{et al.} \cite{nijkamp2023codegen2} trained the \codegen{} on StarCoderData, a programming language dataset developed by BigCode \cite{li2023starcoder}. 
Most recently, Baptiste \textit{et al.} \cite{DBLP:journals/corr/abs-2308-12950} trained \codellama{} starting from a foundation model, Llama 2, which is pre-trained on both general-purpose text and code data.
The results indicate that initializing the model with Llama 2 outperforms the same architecture trained solely on code for a given budget, suggesting that incorporating general-purpose text in the training data can improve the model's performance.

Instruction tuning has been shown to enhance the generalization capability of LLMs on new downstream tasks \cite{DBLP:journals/corr/abs-2303-18223}. 
A study by Yuan \textit{et al.} \cite{yuan2023evaluating} evaluated 10 open-source instructed LLMs on four representative code comprehension and generation tasks, including defect detection, clone detection, assertion generation, and code summarization. 
The study revealed that instructed LLMs perform competitively on code comprehension and generation tasks, occasionally surpassing small state-of-the-art models specifically fine-tuned for each downstream task. 
Another work by Luo \textit{et al.} \cite{luo2023wizardcoder} introduced WizardCoder, which enhances LLMs with complex instruction fine-tuning by adapting the Evol-Instruct method to the domain of code.
The results demonstrate that the fine-tuned model significantly outperforms all other open-source Code LLMs.

\subsection{Security of LLMs}

Pearce \textit{et al.} \cite{pearce2022asleep} systematically investigated the prevalence and conditions under which GitHub Copilot might recommend insecure code. 
The paper used a subset of MITRE's CWE Top 25 list and designed scenarios for Copilot to complete that could potentially introduce a CWE in the generated code. 
The paper created 89 different scenarios for Copilot to complete, resulting in 1,689 programs. 
Of these, approximately 40\% were found to be vulnerable. 
Building on this work, Sandoval \textit{et al.} \cite{DBLP:conf/uss/SandovalPNKGD23} conducted a security-driven user study to assess the impact of LLM-based code assistants on the code quality and security of novice programmers.
They tasked 58 computer science students with implementing a singly-linked list in C, with or without the assistance of a Codex-based LLM.
They discovered that the LLM-assisted group produced functionally better code but also introduced security bugs at a rate no greater than 10\% higher than the control group.
Another work by Khoury \textit{et al.} \cite{DBLP:journals/corr/abs-2304-09655} examined the security of code generated by ChatGPT by asking ChatGPT to generate 21 programs in different programming languages.
They found that ChatGPT often generates insecure code but can also provide useful explanations and suggestions for improvement.

\subsection{Security Weaknesses in IaC Scripts}

While a large body of research focuses on vulnerability detection in traditional source code \cite{fan2020ac,bhandari2021cvefixes,chen2023diversevul,ding2024vulnerability,li2024cleanvul,li2025out,weyssow2025r2vul}, security weaknesses in Infrastructure-as-Code (IaC) scripts represent another important area of study.
IaC has gained popularity for automating network node management.
However, its scripts can contain security weaknesses. 
Research efforts have been directed towards improving IaC script quality by focusing on recognizing these security issues.
Rahman \textit{et al.} \cite{DBLP:journals/tosem/RahmanRPW21} conducted a study to identify security weaknesses in Ansible and Chef scripts. 
They developed a static analysis tool called SLAC to automatically identify six security weaknesses.
Later, Rahman and Williams \cite{DBLP:journals/ieeesp/RahmanW21} identified 67,801 occurrences of security weaknesses, emphasizing the need for more comprehensive quality checkers for IaC code. 
Recently, Saavedra \textit{et al.} \cite{saavedra2023polyglot} introduced GLITCH, a technology-agnostic framework for automated detection of nine major security weaknesses in IaC scripts.
The study demonstrated that GLITCH has higher precision and recall than current state-of-the-art tools. 
These studies underscore the importance of recognizing security weaknesses in IaC scripts to ensure the security and reliability of the deployed infrastructure.
However, the existing tools are limited to recognizing security weaknesses and cannot generate new code.
Previous studies \cite{rahman2019seven,DBLP:journals/tosem/RahmanRPW21,saavedra2023polyglot} conducted a qualitative analysis on 1,726 IaC scripts to recognize and map each security weakness defined by the CWE.
They ultimately identified nine major security weaknesses.
Therefore, in this study, we focus on the same nine major security weaknesses in the state-of-the-art research \cite{rahman2019seven,DBLP:journals/tosem/RahmanRPW21,saavedra2023polyglot} and investigate the capabilities of LLMs to recognize these security weaknesses during the generation and inspection of IaC scripts. 
The definitions of these security weaknesses are as follows:

\paragraph{Admin by Default (CWE-250)}
This weakness arises when default users are granted administrative privileges, potentially violating the principle of least privilege and allowing unauthorized access.

\paragraph{Empty Password (CWE-258)}
This weakness occurs when a password is set to an empty string, constituting a weak password that can be easily guessed.

\paragraph{Hard-Coded Secret (CWE-798)}
This weakness emerges when sensitive information, such as usernames, passwords, and private keys, are exposed as configurations in IaC scripts.
This can reveal the credentials to attackers and compromise the system's security.

\paragraph{Unrestricted IP Address (CWE-284)}
This weakness appears when the address 0.0.0.0 is assigned to a database server or a cloud service/instance, potentially allowing connections from all possible networks and exposing the server or service to potential attacks.

\paragraph{Suspicious Comments (CWE-546)}
This weakness occurs when comments contain information about defects, missing functionality, or system weaknesses, which can reveal weaknesses or mislead other developers.

\paragraph{Use of HTTP Without TLS (CWE-319)}
This weakness arises when HTTP is used without Transport Layer Security (TLS), making the communication between two entities less secure and susceptible to man-in-the-middle attacks.

\paragraph{Use of Weak Cryptography Algorithms (CWE-327)}
This vulnerability arises when cryptographic hash functions with known weaknesses, such as MD4 and SHA-1, are utilized in security-critical applications.
These hash functions are susceptible to collision attacks, which can compromise the integrity of the hashed data.

\paragraph{No Integrity Check (CWE-494)}
This weakness is the recurring pattern of downloading content from the internet without checking the downloaded content using checksums or GPG signatures. 

\paragraph{Missing Default in Case Statement (CWE-478)}
This weakness is the recurring pattern of not handling all input combinations when implementing case conditional logic.

\vspace{-2mm}
\section{Preliminary Base LLM Evaluation}
\label{sec:base_llm}

Given the lack of prior assessments, we first perform a preliminary evaluation for both general-purpose and code-specific LLM models in terms of their capabilities to recognize nine main security weaknesses during the generation and inspection of IaC code.

\vspace{-2mm}
\subsection{Model Choices} 
Our experiments encompass various state-of-the-art LLMs to ensure a comprehensive evaluation of their performance in identifying IaC security weaknesses during IaC code generation and inspection. We have carefully selected four cutting-edge code-specific LLMs, which include \starcoder{} \cite{li2023starcoder}, \codegen{} \cite{nijkamp2023codegen2}, \codellama{} \cite{DBLP:journals/corr/abs-2308-12950}, and \wizardcoder{} \cite{luo2023wizardcoder}. These models have demonstrated exceptional performance in code-related tasks, making them suitable candidates for our study. We utilize the multi-language versions of these LLMs as they support IaC languages, ensuring compatibility with our research objectives.
To provide a more robust comparison, we also include four general-purpose LLMs in our experiments. Two of these models are open-source, namely \vicuna{} \cite{vicuna2023} and \llama{} \cite{touvron2023llama}, while the other two are closed-source, \gpt{} and \gptfour{} \cite{brown2020language}. By incorporating both code-specific and general-purpose LLMs, we aim to gain a deeper understanding of their respective strengths and weaknesses in the context of IaC security.
It is important to note the specific versions of the closed-source models used in our experiments. For \gpt{}, we employ the chatgpt-35-0301 version, while for \gptfour{}, we utilize the gpt-4-0314 version.

\subsection{Evaluating Security Weakness Detection Performance}

To assess security weakness detection performance during the generation and inspection of IaC code, we curate two separate testing datasets.
One dataset is designed to assess security weakness detection during the IaC code generation process, while the other concentrates on detection during IaC code inspection.
To ensure full coverage of the nine IaC security weaknesses in the testing dataset, we employ code repositories that were selected by a previous study \cite{DBLP:conf/kbse/Saavedra022} due to their high quality and ample IaC scripts.
Initially, we randomly choose 30 IaC scripts for each type of security weakness from these code repositories, as identified by GLITCH \cite{DBLP:conf/kbse/Saavedra022}.
In total, we select $30 \cdot 9 = 270$ IaC scripts, which include nine IaC security weaknesses.
Since GLITCH may misclassify samples in certain cases, it is crucial to manually verify the test dataset to establish an accurate ground truth. 
Thus, we examine each identified security weakness and eliminate any false positive samples.
Following this analysis, we obtain a collection of 148 IaC scripts that contain \textbf{229} security weaknesses in total.
Note that, one script could contain multiple security weaknesses, and each security weakness has more than 20 samples among the 229 security weaknesses.
This guarantees both the quality and quantity of the various security weaknesses, as well as the balance of different types of weaknesses.
Our test dataset is comparable to the only similar study \cite{DBLP:journals/corr/abs-2304-14317} that enhances LLMs for secure code generation (not for IaC code generation), which contains 166 security weaknesses.

Subsequently, we utilize \gptfour{} to generate natural language instructions for scripts in the test dataset, with the aim of assessing the ability of LLMs to generate secure IaC code and determining their capacity to generate code from prompts with varying levels of detail.
Varying levels of detail are important as they enable us to evaluate the LLMs' adaptability and effectiveness in different real-world scenarios.
To achieve this, we convert half of the IaC scripts into low-detailed instructions, while the other half is transformed into high-detailed instructions.
These instructions are then paired with their corresponding IaC scripts to form the testing data (example shown in \cref{fig:example_code_generation}), and a sample of 50 is checked manually to make sure the generated instructions are accurate and correct.
Additionally, for IaC security weakness detection during code inspection, we use the scripts in the test dataset to create pairs of original IaC scripts and response IaC scripts with code comments pinpointing security weaknesses (example shown in \cref{fig:example_code_inspection}). 
This evaluates the LLMs' capability to identify security weaknesses during code inspection.
We ensure that there are no duplicated items in the training and testing datasets, which could potentially lead to overfitting, biased performance evaluation, and an inaccurate representation of the model's ability to generalize to unseen data.

\begin{table*}[ht]
    \centering
    \begin{subtable}[t]{0.3\textwidth}
        \centering
        \resizebox{1.105\columnwidth}{!}{
        \begin{tabular}{C{.7cm}C{1.1cm}C{1.75cm}C{.6cm}C{0.75cm}C{0.75cm}C{0.75cm}}
        \hline
        \multirow{2}{*}{\textbf{\makecell[c]{Task}}} & \multirow{2}{*}{\textbf{\makecell[c]{LLM\\Type}}} & \multirow{2}{*}{\makecell[c]{\textbf{Model}}} & \multirow{2}{*}{\makecell[c]{\textbf{Size}}} & \multicolumn{3}{c}{\textbf{Evaluation Metric}} \\
        \cline{5-7}
        & & & & Pre. & Rec. & F1 \\
        \hline
        \parbox[t]{2mm}{\multirow{8}{*}{\rotatebox[origin=c]{90}{Code Generation}}} & \multirow{4}{*}{\makecell[c]{Code\\Specific}} & \codegen{} & 7B & 0.600 & 0.158 & 0.250 \\
         & & \starcoder{} & 15B & 0.000 & 0.000 & \underline{0.000} \\
         & & \wizardcoder{} & 15B & 0.359 & 0.135 & 0.196 \\
         & & \codellama{} & 13B & 0.459 & 0.198 & \textbf{0.276} \\
        \cline{2-7}
         & \multirow{4}{*}{\makecell[c]{General\\Purpose}} & \vicuna{} & 13B & 0.481 & 0.280 & \underline{0.354} \\
         & & \llama{} & 13B & 0.403 & 0.386 & \textbf{0.394} \\
        \cline{3-7}
         & & \gpt{} & - & 0.462 & 0.378 & \underline{0.416} \\
         & & \gptfour{} & - & 0.525 & 0.575 & \textbf{0.549} \\
        \hline
        \parbox[t]{2mm}{\multirow{8}{*}{\rotatebox[origin=c]{90}{Code Inspection}}} & \multirow{4}{*}{\makecell[c]{Code\\Specific}} & \codegen{} & 7B & 0.600 & 0.016 & 0.031 \\
         & & \starcoder{} & 15B & 0.000 & 0.000 & \underline{0.000} \\
         & & \wizardcoder{} & 15B & 0.455 & 0.183 & 0.261 \\
         & & \codellama{} & 13B & 0.633 & 0.199 & \textbf{0.303} \\
        \cline{2-7}
         & \multirow{4}{*}{\makecell[c]{General\\Purpose}} & \vicuna{} & 13B & 0.720 & 0.377 & \textbf{0.495} \\
         & & \llama{} & 13B & 0.588 & 0.298 & \underline{0.396} \\
        \cline{3-7}
         & & \gpt{} & - & 0.664 & 0.476 & \underline{0.555} \\
         & & \gptfour{} & - & 0.673 & 0.529 & \textbf{0.592} \\
        \hline
        \end{tabular}
        }
        \caption{Overall Security}
    \end{subtable}
    \hfill %
    \begin{subtable}[t]{0.69\textwidth}
        \centering
        \resizebox{0.9\columnwidth}{!}{
        \begin{tabular}{C{.7cm}C{1.1cm}C{1.75cm}C{.6cm}C{.75cm}C{.75cm}C{.75cm}C{.75cm}C{.75cm}C{.75cm}C{.8cm}C{.75cm}C{.85cm}C{.75cm}}
        \hline
        \multirow{2}{*}{\textbf{\makecell[c]{Task}}} & \multirow{2}{*}{\textbf{\makecell[c]{LLM\\Type}}} & \multirow{2}{*}{\makecell[c]{\textbf{Model}}} & \multirow{2}{*}{\makecell[c]{\textbf{Size}}} & \multicolumn{9}{c}{\textbf{Type of Security Weakness}} & \multirow{2}{*}{\textbf{\makecell[c]{Avg}}} \\
        \cline{5-13}
        & & & & \scriptsize{AdmDef} & \scriptsize{EmpPw} & \scriptsize{HardCd} & \scriptsize{UnrIP} & \scriptsize{SusCmt} & \scriptsize{NoTLS} & \scriptsize{WeakAlg} & \scriptsize{NoItChk} & \scriptsize{NoDefSw} \\
        \hline
        \parbox[t]{2mm}{\multirow{8}{*}{\rotatebox[origin=c]{90}{Code Generation}}} & \multirow{4}{*}{\makecell[c]{Code\\Specific}} & \codegen{} & 7B & 0.000 & 0.000 & 0.588 & 0.333 & 0.000 & 0.000 & 0.000 & 0.000 & 0.000 & 0.102 \\
        & & \starcoder{} & 15B & 0.000 & 0.000 & 0.000 & 0.000 & 0.000 & 0.000 & 0.000 & 0.000 & 0.000 & \underline{0.000} \\
        & & \wizardcoder{} & 15B & 0.000 & 0.364 & 0.409 & 0.000 & 0.000 & 0.000 & 0.000 & 0.250 & 0.000 & 0.114 \\
        & & \codellama{} & 13B & 0.000 & 0.400 & 0.333 & 0.267 & 0.000 & 0.000 & 0.200 & 0.519 & 0.000 & \textbf{0.191} \\
        \cline{2-14}
        & \multirow{4}{*}{\makecell[c]{General\\Purpose}} & \vicuna{} & 13B & 0.182 & 0.545 & 0.326 & 0.429 & 0.000 & 0.571 & 0.182 & 0.462 & 0.000 & \textbf{0.300} \\
        & & \llama{} & 13B & 0.545 & 0.500 & 0.408 & 0.375 & 0.000 & 0.000 & 0.333 & 0.486 & 0.000 & \underline{0.294} \\
        \cline{3-14}
        & & \gpt{} & - & 0.267 & 0.444 & 0.484 & 0.455 & 0.000 & 0.348 & 0.500 & 0.476 & 0.000 & \underline{0.330} \\
        & & \gptfour{} & - & 0.667 & 0.571 & 0.500 & 0.744 & 0.000 & 0.452 & 0.720 & 0.488 & 0.182 & \textbf{0.480} \\
        \hline
        \parbox[t]{2mm}{\multirow{8}{*}{\rotatebox[origin=c]{90}{Code Inspection}}} & \multirow{4}{*}{\makecell[c]{Code\\Specific}} & \codegen{} & 7B & 0.000 & 0.000 & 0.138 & 0.000 & 0.000 & 0.000 & 0.000 & 0.083 & 0.000 & 0.025 \\
        & & \starcoder{} & 15B & 0.000 & 0.000 & 0.000 & 0.000 & 0.000 & 0.000 & 0.000 & 0.000 & 0.000 & \underline{0.000} \\
        & & \wizardcoder{} & 15B & 0.200 & 0.235 & 0.406 & 0.258 & 0.323 & 0.095 & 0.200 & 0.324 & 0.000 & 0.227 \\
        & & \codellama{} & 13B & 0.316 & 0.353 & 0.464 & 0.412 & 0.074 & 0.111 & 0.400 & 0.364 & 0.000 & \textbf{0.277} \\
        \cline{2-14}
        & \multirow{4}{*}{\makecell[c]{General\\Purpose}} & \vicuna{} & 13B & 0.522 & 0.667 & 0.548 & 0.650 & 0.267 & 0.200 & 0.692 & 0.634 & 0.071 & \textbf{0.472} \\
        & & \llama{} & 13B & 0.381 & 0.333 & 0.528 & 0.650 & 0.000 & 0.174 & 0.643 & 0.424 & 0.000 & \underline{0.348} \\
        \cline{3-14}
        & & \gpt{} & - & 0.640 & 0.455 & 0.514 & 0.714 & 0.556 & 0.533 & 0.741 & 0.711 & 0.000 & 0.540 \\
        & & \gptfour{} & - & 0.750 & 0.783 & 0.484 & 0.941 & 0.207 & 0.619 & 0.815 & 0.583 & 0.000 & \textbf{0.576} \\
        \hline
        \end{tabular}
        }
        \caption{Different Security Weaknesses}
    \end{subtable}
    \label{tb:base_llm_evaluation_comb}
    \caption{Detection of nine security weaknesses in IaC generation and inspection processes.}
    \vspace{-3mm}
\end{table*}

\subsection{Evaluating Syntactical and Functional Correctness}

In order to assess the quality of the generated IaC code scripts, we examine them from two perspectives: syntactical correctness and functional correctness.
For syntactical correctness, we employ basic syntax checkers to determine whether the generated IaC code passes or fails the test, and subsequently calculate the pass rate of these syntax checks.
Evaluating the functional correctness of IaC scripts, however, is considerably more challenging due to the intricate interdependencies between infrastructure components, the dynamic nature of cloud environments, and the absence of standardized testing methodologies. 
A recent study investigated the capacity of LLMs to evaluate generated code \cite{DBLP:journals/corr/abs-2304-14317}, discovering that LLMs such as \gpt{} outperform CodeBERTScore and achieve state-of-the-art performance in evaluating the functional correctness of generated code across various programming languages. 
Consequently, we adopt the framework proposed by \cite{DBLP:journals/corr/abs-2304-14317} and utilize \gptfour{} to assess the functional correctness of the generated IaC code.
The prompt used in the study asks the LLMs to assess the functional correctness concerning execution-based quality of the code snippet on a scale of 0 to 4, with a score of 0 (failing all possible tests) indicating that the code snippet is entirely incorrect and meaningless, and a score of 4 (passing all possible tests) signifying that the code snippet is fully correct and capable of handling all cases.
To align the score with the pass rate range of the syntactical correctness test, we normalize the obtained functional correctness score to also range from 0 to 1.
We further test the approach on a sample of 50 examples from the testing dataset, and the scores provided by \gptfour{} are consistent with our expectations.

\subsection{Prompt Engineering}

To optimize the performance of LLMs in recognizing IaC security weaknesses, we conducted a preliminary study using various prompts to identify the most effective prompt.
We designed three prompts with increasing levels of detail: 1) Write a script following the instructions provided; 2) Write a script following the instructions provided and examine the script and report any security weaknesses; 3) Write a script following the instructions provided, examine the script, report any security weaknesses, and provide definitions of nine major security weaknesses for clarity.
We evaluated the performance of LLMs using the test dataset in recognizing different security weaknesses.
The results indicate that when using the first prompt, LLMs rarely remind users of security weaknesses, which is expected since the prompt does not necessitate security weakness detection. 
When employing the third, more detailed prompt with definitions, all code-specific models struggle to produce valid responses, possibly due to their training primarily on code data rather than natural language, resulting in a limited understanding of natural language.
Consequently, we opted for the second prompt in this study.

\vspace{-3mm}
\subsection{Base LLM Evaluation Results}
\label{sec:base_llm_results}

We then present the results of base LLM experiments on nine main security weaknesses in this section.

\paragraph{Overall Security Weakness Detection}
In Table 1 (a), we present the precision, recall, and F1-score for recognizing security weaknesses in two tasks: IaC code generation and IaC code inspection.
These metrics are reported for various base LLM models, including \codegen{}, \starcoder{}, \wizardcoder{}, \codellama{}, \vicuna{}, \llama{}, \gpt{}, and \gptfour{}.
The highest values are highlighted in bold, while the lowest values are underlined.
Generally, code-specific models exhibit the lowest performance in both tasks, outperformed by open-source general-purpose models (\vicuna{} and \llama{}), which are in turn surpassed by the larger closed-source general-purpose models: \gpt{} and \gptfour{}. 

In particular, \starcoder{} performs the worst, as it fails to identify any security weaknesses during code generation or inspection. 
Other code-specific models achieve F1-scores ranging from 0.196 to 0.276 when recognizing security weaknesses in IaC code generation and 0.031 to 0.303 during IaC code inspection.
Among all code-specific models, \codellama{} improves slightly compared to other code-specific solutions, but is still far from sufficient in identifying security weaknesses, with F1-scores of 0.276 and 0.303 for code generation and inspection, respectively.
This may be attributed to the fact that, unlike most LLMs for code generation (such as CodeGen, InCoder, and StarCoder), which are trained solely on code, \codellama{} is based on a foundation model, Llama2, pretrained on both general-purpose text and code data \cite{DBLP:journals/corr/abs-2308-12950}.

Additionally, we observe that the two open-source general-purpose models, \vicuna{} and \llama{}, exhibit mediocre performance in identifying security weaknesses for both tasks.
Their F1-scores range from 0.354 to 0.394 for code generation and 0.396 to 0.495 for code inspection. 
Specifically, \vicuna{} demonstrates better performance in identifying security weaknesses during code inspection, while \llama{} excels in code generation.
Notably, \gpt{} and \gptfour{} outperform the open-source models, with \gptfour{} surpassing \gpt{} in both tasks by a significant margin.
However, it is crucial to emphasize that even the best-performing models fall significantly short in effectively warning users about security weaknesses.

\paragraph{Breakdown on Different Security Weaknesses}
For a more comprehensive understanding of the capability of different LLMs in recognizing security weaknesses, we present the F1-scores for recognizing various security weaknesses in IaC code generation and inspection tasks in Table 1 (b).
It is evident that some security weaknesses are more easily identified by LLMs compared to others.
Specifically, all LLMs, except \starcoder{}, demonstrate the ability to identify \hardcd{} instances, exhibiting F1-scores within the range of 0.138 to 0.588.
Furthermore, \emppwd{}, \invdip{}, and \noitchk{} are the next most easily detected security weaknesses, with almost two LLMs unable to identify these weaknesses.
Additionally, \notls{}, \weakalg{}, and \admdef{} are the three security weaknesses that can be detected by most general-purpose models and some code-specific models.
In contrast, \suscmt{} and \nodefsw{} are the two most challenging security weaknesses to identify.
Regarding \suscmt{}, it is difficult to recognize during code generation but can be identified during code inspection by most LLMs.
\nodefsw{} can only be detected by \gptfour{} during code generation with a low F1-score of 0.182.

Concerning different models for identifying security weaknesses, it is noticeable that general-purpose models can recognize most security weaknesses with decent accuracy, except for \suscmt{} and \nodefsw{}.
Meanwhile, \codellama{} can also identify the majority of security weakness types, as it is based on the foundation model, \llama{}. 
Interestingly, \llama{} outperforms \codellama{} in both tasks, even though \codellama{} is further trained specifically on code data.
This suggests that training on code data could potentially decrease the performance of LLMs in recognizing security weaknesses.

\paragraph{Syntactical and Functional Correctness}
In \cref{tb:llm_functional_correctness}, we present the results of syntactical correctness and functional correctness for various LLMs in generating IaC code. 
It is worth noting that the prompts have two levels of detail: low and high.
The score ranges from 0 to 1, where 0 signifies completely incorrect, and 1 indicates that the generated code snippet is entirely correct.

As observed, \codegen{} and \starcoder{} exhibit the lowest performance among all LLMs, with syntactical correctness scores ranging from 0.298 to 0.817 and functional correctness scores from 0.420 to 0.613.
Interestingly, these two models demonstrate better performance with low-detailed prompts compared to high-detailed prompts in terms of both syntactical and functional correctness. 
This suggests that \codegen{} and \starcoder{} process abstract prompts more effectively than detailed prompts. 
This might be attributed to the fact that these two models are primarily trained on code data and have limited understanding of natural language.

\begin{table}[htb]
\caption{Comparison between different LLMs in generating syntactically and functionally correct code.}
\label{tb:llm_functional_correctness}
\begin{center}
\resizebox{\columnwidth}{!}{
\begin{tabular}{C{2.2cm}C{.8cm}C{1.cm}C{1.cm}C{1.cm}C{1.cm}}
\hline
\multirow{2}{*}{\makecell[c]{\textbf{Model}}} & \multirow{2}{*}{\makecell[c]{\textbf{Size}}} & \multicolumn{2}{c}{\textbf{Syntax Valid.}} & \multicolumn{2}{c}{\textbf{Functional Corr.}} \\
\cline{3-4} 
\cline{5-6}
& & Low & High & Low & High \\
\hline
\codegen{} & 7B & 0.817 & \underline{0.298} & \underline{0.468} & 0.468 \\
\starcoder{} & 15B & \underline{0.756} & 0.476 & 0.613 & \underline{0.420} \\
\wizardcoder{} & 15B & \textbf{0.915} & 0.786 & \textbf{0.910} & 0.757 \\
\codellama{} & 13B & 0.768 & \textbf{0.881} & 0.733 & \textbf{0.854} \\
\hline
\vicuna{} & 13B & \textbf{0.927} & \textbf{0.952} & \textbf{0.872} & \textbf{0.900} \\
\llama{} & 13B & \underline{0.915} & \underline{0.857} & \underline{0.797} & \underline{0.781} \\
\hline
\gpt{} & - & \underline{0.915} & \underline{0.964} & \underline{0.869} & \underline{0.951} \\
\gptfour{} & - & \textbf{1.000} & \textbf{1.000} & \textbf{0.992} & \textbf{1.000} \\
\hline
\end{tabular}
}
\end{center}
\vspace{-3mm}
\end{table}

In contrast, the other two code-specific models (\wizardcoder{} and \codellama{}) outperform \codegen{} and \starcoder{}, achieving both syntactical and functional correctness scores above 0.7. 
Their performance is on par with the other two open-source general-purpose models, \vicuna{} and \llama{}.
Regarding GPT models, their exceptional size enables them to achieve impressive performance in both syntactical correctness and functional correctness. 
Specifically, the performance of \gpt{} is slightly better than \vicuna{}, while the performance of \gptfour{} is nearly perfect, with the only exception being the functional correctness for low-detailed prompts, which stands at 0.992.

\paragraph{Examples of Nine Prominent IaC Security Weaknesses}
We present examples of responses from LLMs recognizing nine major IaC security weaknesses during the two security weakness detection evaluation tasks mentioned above.
These examples demonstrate that LLMs can recognize and identify these nine prominent security weaknesses to a certain extent.

\paragraph{Example I: CWE-250, Admin by Default} The code containing \admdef{} is shown below, where the \emph{user} is set to be \emph{root}.

\begin{lstlisting}[language=ruby, rulecolor=\color{black}]]
- name: Install MariaDB Galera Cluster
  hosts: ""kolide-db_all""
  become: true
  user: root
\end{lstlisting}

Only \gptfour{} reported this security weakness with the response: \textit{``Running tasks as the `root' user can be a security risk, as it provides full access to the system. It is recommended to use a less privileged user and escalate privileges only when necessary using `become' or `become\_user'.''}

\paragraph{Example II: CWE-258, Empty Password} The code containing \emppwd{} is shown below, where the \emph{login password} is left empty.

\begin{lstlisting}[language=ruby, rulecolor=\color{black}]
- name: set root password
  community.mysql.mysql_user:
    login_unix_socket: ""{{ mysql_socket }}""
    login_user: root
    login_password: """"
\end{lstlisting}

This security weakness is identified by \llama{}, \gpt{}, and \gptfour{}.
\llama{} responds with \textit{``The `login\_password' parameter is set to an empty string, which could potentially allow unauthorized access to the MySQL server.''}

\paragraph{Example III: CWE-798, Hard-Coded Secret} \hardcd{} occurs when sensitive information is hard-coded. In this case, both the username and password are hard-coded.

\begin{lstlisting}[language=ruby, rulecolor=\color{black}]
- name: Create user
  openstack.cloud.identity_user:
     cloud: ""{{ cloud }}""
     state: present
     name: ansible_user
     password: secret
\end{lstlisting}

Apart from \starcoder{} and \codegen{}, all other LLMs reported this security weakness.
For example, \vicuna{} responds with \textit{``The password for the user "ansible\_user" is hardcoded as "secret" in the `openstack.cloud.identity\_user' task. This is a security risk as it means that anyone with access to the playbook can easily guess the password for the user.''}

\paragraph{Example IV: CWE-284, Unrestricted IP Address} This weakness occurs because the unrestricted IP address (i.e., \emph{memcached listen ip}) is assigned, which can be accessed from every possible network.

\begin{lstlisting}[language=ruby, rulecolor=\color{black}]
memcached_listen_ip: ""0.0.0.0""
\end{lstlisting}

This security weakness is identified by \llama{}, \gpt{}, and \gptfour{}.
\llama{} responds with \textit{``The memcached listen IP is set to "0.0.0.0", which means that the memcached server is listening on all interfaces. This can be a security risk, as it allows any incoming connection to the memcached server. It's better to restrict the memcached listen IP to a specific interface or IP address range.''}

\paragraph{Example V: CWE-546, Suspicious Comments} The code comments in this code snippet indicate that naming is non-optimal.

\begin{lstlisting}[language=ruby, rulecolor=\color{black}]
# XXX on windows it's named differently :(
# Best fix: switch to c:/builds/slave on windows
# Other fix: create dirs::builds::moz2_slave and use that 
# in puppet
'C:/builds/moz2_slave':
  ensure => directory;
\end{lstlisting}

This security weakness is only detected by \vicuna{} and \gptfour{}.
\vicuna{} reports the security weakness with the response: \textit{``The use of `XXX` as a comment in the code, which is not secure and should be replaced with a proper fix.''}

\paragraph{Example VI: CWE-319, Use of HTTP Without TLS} This code contains \notls{} as shown below, as it uses the HTTP link to download, which is susceptible to man-in-the-middle attacks.

\begin{lstlisting}[language=ruby, rulecolor=\color{black}]
apt_repository ""apache-cassandra"" do
  uri ""http://www.apache.org/dist/cassandra/debian""
\end{lstlisting}

This security smell is only identified by \gptfour{} with the response: \textit{``The repository URL "http://www.apache.org/dist/cassandra/debian" is not using HTTPS. It is recommended to use HTTPS to ensure secure communication between the client and the server.''}

\paragraph{Example VII: CWE-327, Use of Weak Cryptography Algs} This code uses MD5, which is considered insecure by modern standards. MD5 is vulnerable to collision attacks and is not recommended.

\begin{lstlisting}[language=ruby, rulecolor=\color{black}]
hba_configuration([
  {type: 'host', database: 'all', user: 'all', address: '0.0.0.0/0', method: 'md5'},
])
\end{lstlisting}

This security smell is only identified by \gpt{} and \gptfour{}.
\gptfour{} reports the smell with the response: \textit{``This configuration allows any user to connect to any database from any IP address using the md5 authentication method... consider using a more secure authentication method like scram-sha-256 instead of md5.''}

\paragraph{Example VIII: CWE-494, No Integrity Check} \noitchk{} occurs in this code because \emph{gpgcheck} is set to be \emph{false}, which disables the GPG (GNU Privacy Guard) signature verification for the packages being installed. GPG signatures are a way to ensure that the packages are from a trusted source and have not been tampered with since they were signed.

\begin{lstlisting}[language=ruby, rulecolor=\color{black}]
- name: create repo
  yum_repository:
    name: aap_installer
    description: aap_installer
    baseurl: ""file:///{{ aap_dir }}/bundle/el8/repos""
    gpgcheck: false
\end{lstlisting}

This security smell is identified by \llama{}, \vicuna{}, \gpt{}, and \gptfour{}.
\codellama{} reports this smell with the response: \textit{``The `gpgcheck' parameter is set to `false', which means that the repository is not verified using GPG signatures. This can lead to potential man-in-the-middle attacks, where an attacker can intercept and modify the repository contents before they are downloaded.''}

\paragraph{Example IX: CWE-478, Missing Default in Case Statement} The code containing \nodefsw{} is shown below, as it does not handle all input combinations. Unhandled cases can be exploited in security attacks. If an attacker discovers that certain inputs are not properly handled, they might use these inputs to cause the system to behave in an unintended way.

\begin{lstlisting}[language=ruby, rulecolor=\color{black}]
case node['platform_family']
when 'debian'
  package 'libapache2-mod-wsgi'
when 'rhel', 'fedora', 'arch'
  package 'mod_wsgi' do
    notifies :run, 'execute[generate-module-list]', :immediately
  end
end
\end{lstlisting}

This security smell is only identified by \gptfour{} during code generation.
\gptfour{} reports the security smell while generating the code: \textit{``The script does not handle the case when the platform family is not 'debian', 'rhel', 'fedora', or 'arch'. This might lead to unexpected behavior or security issues if the script is run on an unsupported platform.''}

\paragraph{Examples of Unaware IaC Security Weaknesses}
Previous studies \cite{rahman2019seven,opdebeeck2023control,saavedra2023polyglot} have summarized and developed tools to identify nine major IaC security weaknesses. 
Although the state-of-the-art studies \cite{rahman2019seven,opdebeeck2023control,saavedra2023polyglot} assert that they adhered to the CWE list for identifying IaC security weaknesses, LLMs have also uncovered several unrecognized security weaknesses that could be integrated into existing IaC security weakness detection tools.
A selection of these previously overlooked security weaknesses is provided below.
By incorporating these newly discovered weaknesses, researchers can expand the current list of IaC security weaknesses and enhance the scope and effectiveness of existing state-of-the-art tools.

\paragraph{Example I: CWE-276, Insecure File Permissions} This weakness arises in the code shown below due to the file permission being set to 0755, which is excessively permissive.

\begin{lstlisting}[language=ruby, rulecolor=\color{black}]
node[:opsworks_initial_setup][:bind_mounts][:mounts].each do |dir, source|
  directory source do
    recursive true
    action :create
    mode ""0755""
  end
end
\end{lstlisting}

This security weakness is identified by \codellama{}, \llama{}, and \gptfour{}.
\codellama{} reports the weakness with the response: \textit{``The `directory' resource is used to create a directory with permissions `0755', which is a world-readable and writable directory. This is a security risk, as it allows any user on the system to read and write to the directory.''}

\paragraph{Example II: CWE-269, Privilege Escalation} This vulnerability occurs when a software system fails to properly manage the privileges assigned to it, enabling users or processes to perform actions with more privileges than they should have. In this instance, \emph{sudo} is set to \emph{yes}.

\begin{lstlisting}[language=ruby, rulecolor=\color{black}]
- name: Install tempest from rpm
  hosts: tester
  gather_facts: no
  sudo: yes
\end{lstlisting}

This security weakness is identified by all LLMs except \starcoder{} and \codegen{}.
\vicuna{} reports the security weakness with \textit{``The `sudo: yes' directive in the `hosts' section indicates that the task is running with elevated privileges. This can be a security risk if the task is not properly secured and could be exploited by an attacker to gain access to sensitive information or perform unauthorized actions.''}

\paragraph{Example III: CWE-20, Lack of Input Validation} This vulnerability occurs when the software does not validate or incorrectly validates input that can affect the control flow or data flow of a program.

\begin{lstlisting}[language=ruby, rulecolor=\color{black}]
package { 'iTerm':
  ensure   => $ensure,
  flavor   => 'zip',
  provider => 'compressed_app',
  source   => ""http://www.iterm2.com/downloads/stable/iTerm2-${version}.zip""
}
\end{lstlisting}

This security weakness is identified by \wizardcoder{}, \codellama{}, \gpt{}, and \gptfour{}.
\wizardcoder{} reports the security weakness: \textit{``...the `ensure` parameter is not validated, which could allow an attacker to bypass the package management system and install unauthorized software.''}

\paragraph{Example IV: CWE-77, Potential Command Injection} This vulnerability occurs when an application does not properly sanitize external input, potentially allowing an attacker to execute unintended commands.

\begin{lstlisting}[language=ruby, rulecolor=\color{black}]
bash 'install npm - package manager for node' do
  cwd '/usr/local/src'
  user 'root'
  code <<-EOH
    mkdir -p npm-v#{node['nodejs']['npm']} && \
    cd npm-v#{node['nodejs']['npm']}
    curl -L #{npm_src_url} | tar xzf - --strip-components=1 && \
    make uninstall dev
  EOH
end
\end{lstlisting}

This security weakness is identified by \llama{}, \codellama{}, \gpt{}, and \gptfour{}.
\llama{} reports the security weakness with \textit{``The code uses `code' to execute a shell command. This could potentially expose the system to command injection attacks.''}

\section{\tool{}: Approach}
\label{sec:approach}

In this section, we present our approach to address the limitations of base LLMs in recognizing security weaknesses while generating or inspecting IaC code.

\subsection{Training}

Fine-tuning LLMs with billions of parameters is computationally demanding. To expedite training and save resources, parameter-efficient methods like adapter tuning \cite{houlsby2019parameter}, prefix tuning \cite{li2021prefix}, prompt tuning \cite{lester2021power}, and LoRA \cite{DBLP:conf/iclr/HuSWALWWC22} are crucial. We focus on LoRA, a popular method for efficient LLM tuning. LoRA, or Low-Rank Adaptation, uses a low-rank matrix to model input embeddings and pre-trained model parameters interactions, reducing trainable parameters and enabling faster adaptation. LoRA factorizes the weight matrix into two smaller matrices, decreasing fine-tuning parameters while retaining the LLM's expressive power.
LoRA offers several advantages, including low-rank matrix factorization for compact parameter representation, faster training, and reduced memory usage. Its parameter-efficient fine-tuning enables better generalization to new tasks and robust knowledge transfer. Additionally, LoRA's simplicity and compatibility with existing LLM architectures make it an appealing, resource-efficient option for fine-tuning LLMs.

We configure LoRA using three key parameters: the low-rank matrix rank (\emph{R}) is set to 16, which determines the dimensionality of the low-rank approximation; the scaling factor (\emph{alpha}) is set to 32, which controls the balance between the original model and the low-rank adaptation; and the regularization parameter (\emph{dropout}) is set to 0.05, which helps prevent overfitting by randomly dropping out a proportion of the model's weights during training.
Furthermore, we employ 8-bit quantization, a technique that compresses the model parameter space by representing each parameter with only 8 bits, effectively reducing GPU memory consumption during fine-tuning.
To ensure the availability of long training sequences, we set the maximum token length to 1024. 
Regarding the number of epochs, we set it to 3 based on experimental findings. 
Through experiments, we observed that there is no significant performance improvement beyond 3 epochs, indicating that the model converges rapidly and additional training would not yield further benefits.
We also maintain the learning rate consistent with LLM's pretraining learning rate to ensure a smooth transition from pretraining to fine-tuning.
Our experiments were conducted on NVIDIA A100 GPUs. 
For the largest LLMs (15B) fine-tuned in our study, our training approach is relatively cost-effective, requiring less than 10 hours per epoch and under 30GB of GPU memory. 
This is in stark contrast to LLM pretraining, which necessitates GPU clusters and extensive time investments ranging from weeks to months \cite{li2023starcoder,nijkamp2023codegen2,DBLP:journals/corr/abs-2308-12950}.

\subsection{\tool{} Dataset}

We introduce our dataset for fine-tuning LLM models to enhance IaC security weakness detection while generating and inspecting IaC code.
The quality of the training dataset is crucial for achieving accurate models and ensuring real-world effectiveness.
To construct the training dataset for instruction tuning, we adhere to the guidelines provided by Zhao \textit{et al.} \cite{DBLP:journals/corr/abs-2303-18223}. 
We utilize code repositories chosen by a previous study \cite{DBLP:conf/kbse/Saavedra022} due to their high quality, sufficient IaC scripts, active development, and appropriate scale for this research.
Specifically, this dataset comprises Ansible, Chef, and Puppet scripts.
The Ansible dataset was built from 681 repositories and contains 108K Ansible scripts; the Chef dataset was assembled from 439 repositories and contains 70K Chef scripts, and the Puppet dataset was constructed from 293 repositories and contains 17K Puppet scripts.
In total, the three IaC datasets consist of 1413 repositories with 196K IaC scripts.
Our focus is on enhancing security weakness detection capabilities while generating and inspecting IaC code, so we create our dataset with two components: code generation and code inspection.

\paragraph{Code Generation Data} To improve security weakness detection performance during code generation, we require pairs of natural language descriptions and corresponding IaC code. 
As mentioned in \cite{DBLP:journals/corr/abs-2303-18223}, although real-world instructions from human users are more suitable for fine-tuning LLMs, collecting them at a large scale is challenging. 
It is also essential to have a sufficient number of instructions in the training dataset. 
As a result, most existing research primarily uses synthetic instructions generated by LLMs \cite{DBLP:journals/corr/abs-2303-18223}. 
We first employ the state-of-the-art tool GLITCH \cite{DBLP:conf/kbse/Saavedra022} to analyze the aforementioned IaC dataset and recognize all nine supported IaC security weaknesses.
We then extract the IaC scripts containing security weaknesses and use \gptfour{} to convert the IaC code into two levels (low detailed and high detailed) of natural language instructions.
Having both low and high detailed levels of instructions is crucial for enhancing instruction complexity, which can improve the quality of the training dataset \cite{DBLP:journals/corr/abs-2303-18223}.

\begin{figure}[htb]
\begin{lstlisting}[language=C]
(*@\textbf{Low-Detailed Level Instruction:}@*)
Creates a new user on all target hosts using Ansible, adds a user named 'ansible' with the password '@ns1bl3', and ensures the user is present.
\end{lstlisting}
\begin{lstlisting}[language=C]
(*@\textbf{High-Detailed Level Instruction:}@*)
1. Create a new Ansible playbook file.
2. Add a play with the name "Add a user".
3. Set the target hosts to "all".
4. Disable gathering facts by setting "gather_facts" to "false".
5. Add a task with the name "Add User".
6. Use the "win_user" module.
7. Set the "name" parameter to "ansible".
8. Set the "password" parameter to "@ns1bl3".
9. Set the "state" parameter to "present".
\end{lstlisting}
\begin{lstlisting}[language=ruby, numbers=left, numberstyle=\linecolor{10}{color-f}]
(*@\textbf{IaC Script:}@*)
---
- name: Add a user
  hosts: all
  gather_facts: false
  tasks:
    - name: Add User
      win_user:
        name: ansible
        # Security smell! Hard-coded secret...
        password: "@ns1bl3"
        state: present
\end{lstlisting}
\vspace{-3mm}
\caption{Example of Code Generation Data.}
\label{fig:example_code_generation}
\vspace{-3mm}
\end{figure}

\begin{figure}[htb]
\begin{lstlisting}[language=ruby]
(*@\textbf{Input IaC Script:}@*)
---
class gnome::install{
  case $operatingsystem {
    'Debian': {
      package { ['task-gnome-desktop']:
        ensure => 'installed',
      }
    }
  }
}
\end{lstlisting}
\begin{lstlisting}[language=ruby, numbers=left, numberstyle=\linecolor{5}{color-f}]
(*@\textbf{Response IaC Script:}@*)
---
class gnome::install{
  case $operatingsystem {
    # Security smell! Missing default in case statement
    'Debian': {
      package { ['task-gnome-desktop']:
        ensure => 'installed',
      }
    }
  }
}
\end{lstlisting}
\vspace{-3mm}
\caption{Example of Code Inspection Data.}
\label{fig:example_code_inspection}
\vspace{-3mm}
\end{figure}

An example of a code generation data pair is shown in \cref{fig:example_code_generation}. 
The low-detailed level instruction is displayed at the top, providing a brief description of the task that captures its essence while omitting most detailed information.
In contrast, the high-detailed level instruction is shown in the middle, offering a step-by-step guide to implementing the required features using pseudo-code or natural language.
To achieve a balanced dataset with varying levels of detail in instructions \cite{DBLP:journals/corr/abs-2303-18223}, we include 11K IaC scripts with security weaknesses, converting 5.5K IaC scripts to low-detailed instructions and the remaining 5.5K scripts to high-detailed instructions.
We also add code comments (highlighted in green) in the IaC script to pinpoint the location and names of security weaknesses using the security weakness information extracted by GLITCH \cite{DBLP:conf/kbse/Saavedra022}. 
This approach helps LLMs learn to identify security weaknesses during instruction fine-tuning, improving both security weakness detection and IaC code generation capabilities.

\paragraph{Code Inspection Data} To enhance security weakness detection performance during code inspection, we also use the previously mentioned 11K IaC scripts with security weaknesses to curate this dataset. 
An example of an input IaC script and response IaC script pair is shown in \cref{fig:example_code_inspection}. 
Similarly, we add code comments (highlighted in green) in the IaC script to pinpoint the location and names of security weaknesses using the security weakness information extracted by GLITCH \cite{DBLP:conf/kbse/Saavedra022}.
This aims to improve security weakness detection during code inspection. 
In total, we create an instruction fine-tuning dataset comprising 22K data points, with 11K for code generation (example in \cref{fig:example_code_generation}) and 11K for code inspection (example in \cref{fig:example_code_inspection}).

\section{\tool{}: Use Cases}

We discuss practical use cases of \tool{}: recognizing IaC security weaknesses during code generation and inspection.

\paragraph{Generating Secure IaC Code with LLMs}
As developers increasingly rely on LLMs to generate IaC scripts, it is crucial to ensure that the generated code is secure and free from weaknesses.
\tool{} can be employed to enhance the ability of LLMs to recognize IaC security weaknesses during the code generation process, particularly for IaC scripts where most security weaknesses require additional information to be resolved (see example in \cref{fig:example}).
By fine-tuning LLMs with \tool{}, developers can benefit from improved security in the generated IaC code, reducing the risk of potential security breaches and the need for extensive manual inspection.
This use case enables developers to maintain high levels of security while taking advantage of the efficiency and convenience offered by LLMs for code generation.

\paragraph{Recognizing IaC Security Weaknesses with LLMs}
In addition to generating secure IaC code, developers may also use LLMs to inspect existing IaC scripts for potential security weaknesses.
\tool{} can be utilized to improve the performance of LLMs in recognizing IaC security weaknesses during code inspection.
By fine-tuning LLMs with \tool{}, developers can rely on these models to identify and report security weaknesses in IaC scripts more accurately and efficiently.
This use case allows developers to leverage the power of LLMs for recognizing security issues in IaC scripts, streamlining the inspection process, and ensuring a higher level of security in their infrastructure management.

\section{\tool{}: Evaluation}

We assess the performance of LLMs fine-tuned with \tool{} on nine IaC security weaknesses. 
We evaluate the effectiveness of fine-tuning in the following aspects:

\begin{itemize}
    \item \tool{} significantly improves performance in recognizing security weaknesses during the generation and inspection of IaC code (\cref{sec:main_experiments}).
    \item Dataset curation plays a crucial role in enabling \tool{} to achieve optimal performance (\cref{sec:ablation_studies}).
    \item \tool{} can generalize knowledge across IaC languages (\cref{sec:generalizability}).
\end{itemize}

\subsection{\tool{} Main Experiments}
\label{sec:main_experiments}

In this section, we present the results of evaluating \tool{} for IaC code generation and inspection on nine main security weaknesses.
We also compare its performance with its base LLMs and \gptfour{}.
We use consistent color notations that represent base LLMs as \basebox{}, \tool{} as \gensiacbox{}, and GPT4 as \gptbox{} in this section.

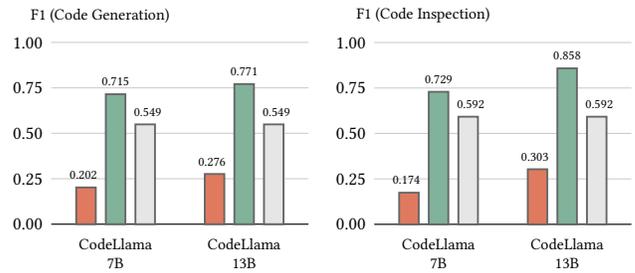
\begin{figure}[htb]
    \begin{minipage}{0.235\textwidth}
        \centering
        \begin{tikzpicture}
            \centering
            \begin{axis}[
                height=4cm, width=5cm,
                /pgf/bar width=0.26cm,
                xmin=-0.2, xmax=0.8,
                axis x line*=bottom, axis y line*=left, enlarge x limits=true,
                xtick={0, 0.6},
                xticklabel style={yshift=-0.8mm, font=\scriptsize, align=center},
                ybar=3.8pt, clip=false,
                ymin=0, ymax=1, ytick={0, 0.25, 0.5, 0.75, 1.0}, yticklabels={0.00, 0.25, 0.50, 0.75, 1.00},
                ymajorgrids, major grid style={draw=black!20}, tick align=inside,
                yticklabel style={font=\footnotesize}, tickwidth=0pt,
                y axis line style={opacity=0},
                ylabel={\scriptsize F1 (Code Generation)},
                y label style={at={(0.55, 1.15)}, rotate=-90},
                xticklabels={\shortstack[c]{\codellama{}\\7B}, \shortstack[c]{\codellama{}\\13B}},
            ]
            
            \addplot [draw=light-gray, line width=0.7pt, fill=color-e, error bars/.cd, y dir=both, y explicit, error bar style={draw=black}] coordinates {
                (0, 0.202)
                (0.6, 0.276)
            };
            \node[above] at ($(axis cs:-0.15, 0.202)$) {\tiny 0.202};
            \node[above] at ($(axis cs:0.45, 0.276)$) {\tiny 0.276};
    
            \addplot [draw=light-gray, line width=0.7pt, fill=color-b, error bars/.cd, y dir=both, y explicit, error bar style={draw=black}] coordinates {
                (0, 0.715)
                (0.6, 0.771)
            };
            \node[above] at ($(axis cs:0, 0.715)$) {\tiny 0.715};
            \node[above] at ($(axis cs:0.6, 0.771)$) {\tiny 0.771};
            
            \addplot [draw=light-gray, line width=0.7pt, fill=gray, error bars/.cd, y dir=both, y explicit, error bar style={draw=black}] coordinates {
                (0, 0.549)
                (0.6, 0.549)
            };
            \node[above] at ($(axis cs:0.15, 0.549)$) {\tiny 0.549};
            \node[above] at ($(axis cs:0.75, 0.549)$) {\tiny 0.549};
            \end{axis}
        \end{tikzpicture}
    \end{minipage}
    \hfill
    \begin{minipage}{0.235\textwidth}
        \centering
        \begin{tikzpicture}
            \centering
            \begin{axis}[
                height=4cm, width=5cm,
                /pgf/bar width=0.26cm,
                xmin=-0.2, xmax=0.8,
                axis x line*=bottom, axis y line*=left, enlarge x limits=true,
                xtick={0, 0.6},
                xticklabel style={yshift=-0.8mm, font=\scriptsize, align=center},
                ybar=3.8pt, clip=false,
                ymin=0, ymax=1, ytick={0, 0.25, 0.5, 0.75, 1.0}, yticklabels={0.00, 0.25, 0.50, 0.75, 1.00},
                ymajorgrids, major grid style={draw=black!20}, tick align=inside,
                yticklabel style={font=\footnotesize}, tickwidth=0pt,
                y axis line style={opacity=0},
                ylabel={\scriptsize F1 (Code Inspection)},
                y label style={at={(0.55, 1.15)}, rotate=-90},
                xticklabels={\shortstack[c]{\codellama{}\\7B}, \shortstack[c]{\codellama{}\\13B}},
            ]
    
            \addplot [draw=light-gray, line width=0.7pt, fill=color-e, error bars/.cd, y dir=both, y explicit, error bar style={draw=black}] coordinates {
                (0, 0.174)
                (0.6, 0.303)
            };
            \node[above] at ($(axis cs:-0.15, 0.174)$) {\tiny 0.174};
            \node[above] at ($(axis cs:0.45, 0.303)$) {\tiny 0.303};
    
            \addplot [draw=light-gray, line width=0.7pt, fill=color-b, error bars/.cd, y dir=both, y explicit, error bar style={draw=black}] coordinates {
                (0, 0.729)
                (0.6, 0.858)
            };
            \node[above] at ($(axis cs:0, 0.729)$) {\tiny 0.729};
            \node[above] at ($(axis cs:0.6, 0.858)$) {\tiny 0.858};
            
            \addplot [draw=light-gray, line width=0.7pt, fill=gray, error bars/.cd, y dir=both, y explicit, error bar style={draw=black}] coordinates {
                (0, 0.592)
                (0.6, 0.592)
            };
            \node[above] at ($(axis cs:0.15, 0.592)$) {\tiny 0.592};
            \node[above] at ($(axis cs:0.75, 0.592)$) {\tiny 0.592};
            \end{axis}
        \end{tikzpicture}
    \end{minipage}
    \caption{Overall F1-score for nine security weaknesses detection.}
    \label{fig:gensiac_overall_f1}
    \vspace{-3mm}
\end{figure}

\paragraph{Overall Security Weakness Detection}
In this section, we present the overall F1-score for \codellama{} models on the detection of nine primary security weaknesses and compare their performance with \gptfour{}, as shown in \cref{fig:gensiac_overall_f1}. 
It is evident that \tool{} significantly enhances the performance of the base model without fine-tuning, achieving an F1-score of over 0.7.
In particular, \codellama{}-7B improves security weakness detection from 0.202 and 0.174 to 0.715 and 0.729 during code generation and inspection phases, respectively.
For \codellama{}-13B, the performance is further enhanced, with F1-scores increasing from 0.715 and 0.729 to 0.771 and 0.858.

When comparing the performance of \codellama{}-13B fine-tuned with \tool{} to \gptfour{}, we observe a substantial improvement.
The F1-scores increase from nearly half of \gptfour{}'s accuracy to significantly higher values, with 0.771 vs. 0.549 and 0.858 vs. 0.592 during code generation and inspection, respectively. 

\begin{figure*}[t]
    \begin{minipage}{\textwidth}
        \centering
        \begin{tikzpicture}
            \centering
            \begin{axis}[
                height=4cm, width=18.2cm,
                /pgf/bar width=0.26cm,
                xmin=0.2, xmax=7.8,
                axis x line*=bottom, axis y line*=left, enlarge x limits=true,
                xtick={0, 1, 2, 3, 4, 5, 6, 7, 8},
                xticklabel style={yshift=-0.8mm, font=\scriptsize, align=center},
                ybar=3.8pt, clip=false,
                ymin=0, ymax=1, ytick={0, 0.25, 0.5, 0.75, 1.0}, yticklabels={0.00, 0.25, 0.50, 0.75, 1.00},
                ymajorgrids, major grid style={draw=black!20}, tick align=inside,
                yticklabel style={font=\footnotesize}, tickwidth=0pt,
                y axis line style={opacity=0},
                ylabel={\scriptsize F1 (Code Generation)},
                y label style={at={(0.118, 1.15)}, rotate=-90},
                xticklabels={\shortstack[c]{AdmDef\\CWE-250}, \shortstack[c]{EmpPwd\\CWE-258}, \shortstack[c]{HardCd\\CWE-798}, \shortstack[c]{UnrIP\\CWE-284}, \shortstack[c]{SusCmt\\CWE-546}, \shortstack[c]{NoTLS\\CWE-319}, \shortstack[c]{WeakAlg\\CWE-327}, \shortstack[c]{NoItChk\\CWE-494}, \shortstack[c]{NoDefSw\\CWE-478}},
            ]
            
            \addplot [draw=light-gray, line width=0.7pt, fill=color-e, error bars/.cd, y dir=both, y explicit, error bar style={draw=black}] coordinates {
                (0, 0.000)
                (1, 0.400)
                (2, 0.333)
                (3, 0.267)
                (4, 0.000)
                (5, 0.000)
                (6, 0.200)
                (7, 0.519)
                (8, 0.000)
            };
            \node[above] at ($(axis cs:-0.25, 0.000)$) {\tiny 0.000};
            \node[above] at ($(axis cs:0.75, 0.400)$) {\tiny 0.400};
            \node[above] at ($(axis cs:1.75, 0.333)$) {\tiny 0.333};
            \node[above] at ($(axis cs:2.75, 0.267)$) {\tiny 0.267};
            \node[above] at ($(axis cs:3.75, 0.000)$) {\tiny 0.000};
            \node[above] at ($(axis cs:4.75, 0.000)$) {\tiny 0.000};
            \node[above] at ($(axis cs:5.75, 0.200)$) {\tiny 0.200};
            \node[above] at ($(axis cs:6.75, 0.519)$) {\tiny 0.519};
            \node[above] at ($(axis cs:7.75, 0.000)$) {\tiny 0.000};
    
            \addplot [draw=light-gray, line width=0.7pt, fill=color-b, error bars/.cd, y dir=both, y explicit, error bar style={draw=black}] coordinates {
                (0, 0.688)
                (1, 0.706)
                (2, 0.827)
                (3, 0.552)
                (4, 0.952)
                (5, 0.636)
                (6, 0.522)
                (7, 0.647)
                (8, 0.885)
            };
            \node[above] at ($(axis cs:0, 0.688)$) {\tiny 0.688};
            \node[above] at ($(axis cs:1, 0.706)$) {\tiny 0.706};
            \node[above] at ($(axis cs:2, 0.827)$) {\tiny 0.827};
            \node[above] at ($(axis cs:3, 0.552)$) {\tiny 0.552};
            \node[above] at ($(axis cs:4, 0.952)$) {\tiny 0.952};
            \node[above] at ($(axis cs:5, 0.636)$) {\tiny 0.636};
            \node[above] at ($(axis cs:6, 0.522)$) {\tiny 0.522};
            \node[above] at ($(axis cs:7, 0.647)$) {\tiny 0.647};
            \node[above] at ($(axis cs:8, 0.885)$) {\tiny 0.885};
            
            \addplot [draw=light-gray, line width=0.7pt, fill=gray, error bars/.cd, y dir=both, y explicit, error bar style={draw=black}] coordinates {
                (0, 0.667)
                (1, 0.571)
                (2, 0.500)
                (3, 0.744)
                (4, 0.000)
                (5, 0.452)
                (6, 0.720)
                (7, 0.488)
                (8, 0.182)
            };
            \node[above] at ($(axis cs:0.25, 0.667)$) {\tiny 0.667};
            \node[above] at ($(axis cs:1.25, 0.571)$) {\tiny 0.571};
            \node[above] at ($(axis cs:2.25, 0.500)$) {\tiny 0.500};
            \node[above] at ($(axis cs:3.25, 0.744)$) {\tiny 0.744};
            \node[above] at ($(axis cs:4.25, 0.000)$) {\tiny 0.000};
            \node[above] at ($(axis cs:5.25, 0.452)$) {\tiny 0.452};
            \node[above] at ($(axis cs:6.25, 0.720)$) {\tiny 0.720};
            \node[above] at ($(axis cs:7.25, 0.488)$) {\tiny 0.488};
            \node[above] at ($(axis cs:8.25, 0.182)$) {\tiny 0.182};
            \end{axis}
        \end{tikzpicture}
    \end{minipage}
    \hfill
    \begin{minipage}{\textwidth}
        \centering
        \begin{tikzpicture}
            \centering
            \begin{axis}[
                height=4cm, width=18.2cm,
                /pgf/bar width=0.26cm,
                xmin=0.2, xmax=7.8,
                axis x line*=bottom, axis y line*=left, enlarge x limits=true,
                xtick={0, 1, 2, 3, 4, 5, 6, 7, 8},
                xticklabel style={yshift=-0.8mm, font=\scriptsize, align=center},
                ybar=3.8pt, clip=false,
                ymin=0, ymax=1, ytick={0, 0.25, 0.5, 0.75, 1.0}, yticklabels={0.00, 0.25, 0.50, 0.75, 1.00},
                ymajorgrids, major grid style={draw=black!20}, tick align=inside,
                yticklabel style={font=\footnotesize}, tickwidth=0pt,
                y axis line style={opacity=0},
                ylabel={\scriptsize F1 (Code Inspection)},
                y label style={at={(0.115, 1.18)}, rotate=-90},
                xticklabels={\shortstack[c]{AdmDef\\CWE-250}, \shortstack[c]{EmpPwd\\CWE-258}, \shortstack[c]{HardCd\\CWE-798}, \shortstack[c]{UnrIP\\CWE-284}, \shortstack[c]{SusCmt\\CWE-546}, \shortstack[c]{NoTLS\\CWE-319}, \shortstack[c]{WeakAlg\\CWE-327}, \shortstack[c]{NoItChk\\CWE-494}, \shortstack[c]{NoDefSw\\CWE-478}},
            ]
            
            \addplot [draw=light-gray, line width=0.7pt, fill=color-e, error bars/.cd, y dir=both, y explicit, error bar style={draw=black}] coordinates {
                (0, 0.316)
                (1, 0.353)
                (2, 0.464)
                (3, 0.412)
                (4, 0.074)
                (5, 0.111)
                (6, 0.400)
                (7, 0.364)
                (8, 0.000)
            };
            \node[above] at ($(axis cs:-0.25, 0.316)$) {\tiny 0.316};
            \node[above] at ($(axis cs:0.75, 0.353)$) {\tiny 0.353};
            \node[above] at ($(axis cs:1.75, 0.464)$) {\tiny 0.464};
            \node[above] at ($(axis cs:2.75, 0.412)$) {\tiny 0.412};
            \node[above] at ($(axis cs:3.75, 0.074)$) {\tiny 0.074};
            \node[above] at ($(axis cs:4.75, 0.111)$) {\tiny 0.111};
            \node[above] at ($(axis cs:5.75, 0.400)$) {\tiny 0.400};
            \node[above] at ($(axis cs:6.75, 0.364)$) {\tiny 0.364};
            \node[above] at ($(axis cs:7.75, 0.000)$) {\tiny 0.000};
    
            \addplot [draw=light-gray, line width=0.7pt, fill=color-b, error bars/.cd, y dir=both, y explicit, error bar style={draw=black}] coordinates {
                (0, 1.000)
                (1, 0.750)
                (2, 0.774)
                (3, 0.875)
                (4, 0.842)
                (5, 0.909)
                (6, 0.815)
                (7, 0.821)
                (8, 0.941)
            };
            \node[above] at ($(axis cs:0, 1.000)$) {\tiny 1.000};
            \node[above] at ($(axis cs:1, 0.750)$) {\tiny 0.750};
            \node[above] at ($(axis cs:2, 0.774)$) {\tiny 0.774};
            \node[above] at ($(axis cs:3, 0.875)$) {\tiny 0.875};
            \node[above] at ($(axis cs:4, 0.842)$) {\tiny 0.842};
            \node[above] at ($(axis cs:5, 0.909)$) {\tiny 0.909};
            \node[above] at ($(axis cs:6, 0.815)$) {\tiny 0.815};
            \node[above] at ($(axis cs:7, 0.821)$) {\tiny 0.821};
            \node[above] at ($(axis cs:8, 0.941)$) {\tiny 0.941};
            
            \addplot [draw=light-gray, line width=0.7pt, fill=gray, error bars/.cd, y dir=both, y explicit, error bar style={draw=black}] coordinates {
                (0, 0.750)
                (1, 0.783)
                (2, 0.484)
                (3, 0.941)
                (4, 0.207)
                (5, 0.619)
                (6, 0.815)
                (7, 0.583)
                (8, 0.000)
            };
            \node[above] at ($(axis cs:0.25, 0.750)$) {\tiny 0.750};
            \node[above] at ($(axis cs:1.25, 0.783)$) {\tiny 0.783};
            \node[above] at ($(axis cs:2.25, 0.484)$) {\tiny 0.484};
            \node[above] at ($(axis cs:3.25, 0.941)$) {\tiny 0.941};
            \node[above] at ($(axis cs:4.25, 0.207)$) {\tiny 0.207};
            \node[above] at ($(axis cs:5.25, 0.619)$) {\tiny 0.619};
            \node[above] at ($(axis cs:6.25, 0.815)$) {\tiny 0.815};
            \node[above] at ($(axis cs:7.25, 0.583)$) {\tiny 0.583};
            \node[above] at ($(axis cs:8.25, 0.000)$) {\tiny 0.000};
            \end{axis}
        \end{tikzpicture}
    \end{minipage}
    \caption{F1-score for the detection of nine security weaknesses.}
    \label{fig:gensiac_smell_individual_f1}
\end{figure*}
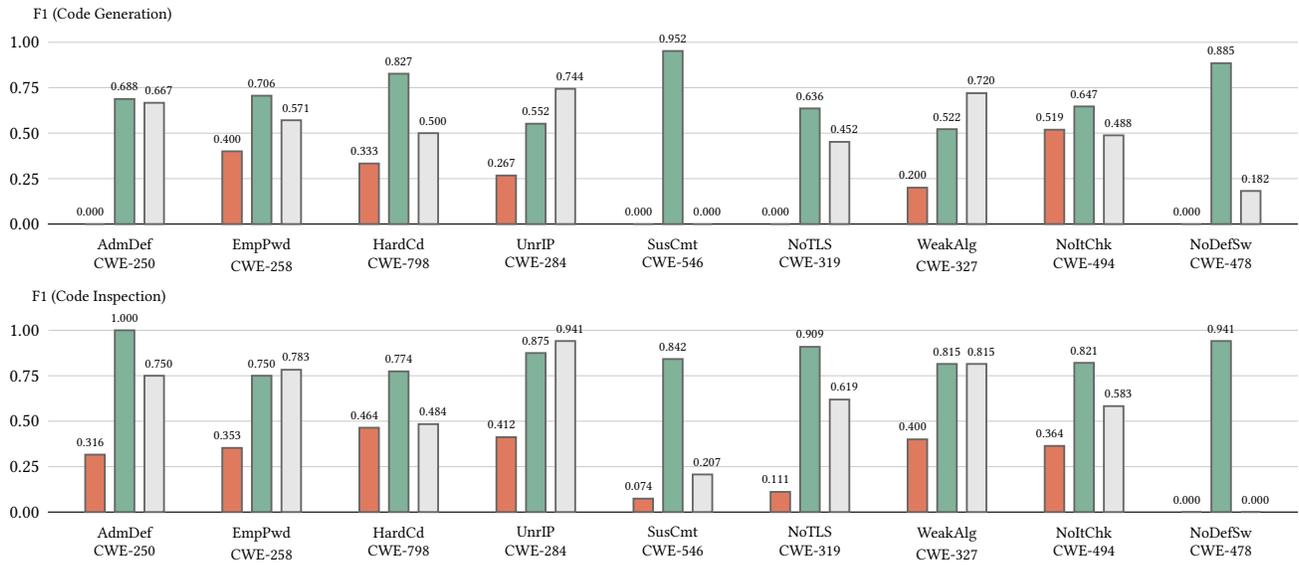

\paragraph{Breakdown on Different Security Weaknesses}
To gain a deeper understanding of \tool{}'s capability in identifying different security weaknesses, we present the F1-score for nine distinct security weaknesses during code generation and inspection in \cref{fig:gensiac_smell_individual_f1}.
It can be observed that \tool{} consistently improves the F1-score for all security weaknesses across both tasks when compared to the base \codellama{}.
In particular, \tool{} demonstrates substantial improvements (over 100\% in terms of F1-score) in recognizing \admdef{}, \invdip{}, \suscmt{}, \notls{}, \weakalg{}, and \nodefsw{}. 
The F1-score improvements are 0.688, 0.285, 0.952, 0.636, 0.322, and 0.885 during code generation, and 0.684, 0.463, 0.768, 0.798, 0.415, and 0.941 during code inspection, respectively.

Furthermore, when comparing \tool{} to \gptfour{}, \tool{} almost always exhibits increased or equivalent performance in security weakness detection. The exceptions are \invdip{} and \weakalg{}, where \tool{} experiences a decrease of 0.192 and 0.198 during code generation. 
However, for \hardcd{}, \suscmt{}, \notls{}, \noitchk{}, and \nodefsw{}, \tool{} shows significant improvement over \gptfour{}.

\begin{figure}[htb]
    \begin{minipage}{0.235\textwidth}
        \centering
        \begin{tikzpicture}
            \centering
            \begin{axis}[
                height=4cm, width=5cm,
                /pgf/bar width=0.26cm,
                xmin=-0.2, xmax=0.8,
                axis x line*=bottom, axis y line*=left, enlarge x limits=true,
                xtick={0, 0.6},
                xticklabel style={yshift=-0.8mm, font=\scriptsize, align=center},
                ybar=3.8pt, clip=false,
                ymin=0, ymax=1, ytick={0, 0.25, 0.5, 0.75, 1.0}, yticklabels={0.00, 0.25, 0.50, 0.75, 1.00},
                ymajorgrids, major grid style={draw=black!20}, tick align=inside,
                yticklabel style={font=\footnotesize}, tickwidth=0pt,
                y axis line style={opacity=0},
                ylabel={\scriptsize Pass Rate (Syntax Valid.)},
                y label style={at={(0.55, 1.2)}, rotate=-90},
                xticklabels={\shortstack[c]{Low-Detailed\\Prompt}, \shortstack[c]{High-Detailed\\Prompt}},
            ]
            
            \addplot [draw=light-gray, line width=0.7pt, fill=color-e, error bars/.cd, y dir=both, y explicit, error bar style={draw=black}] coordinates {
                (0, 0.890)
                (0.6, 0.964)
            };
            \node[above] at ($(axis cs:-0.15, 0.890)$) {\tiny 0.890};
            \node[above] at ($(axis cs:0.45, 0.964)$) {\tiny 0.964};
    
            \addplot [draw=light-gray, line width=0.7pt, fill=color-b, error bars/.cd, y dir=both, y explicit, error bar style={draw=black}] coordinates {
                (0, 1.000)
                (0.6, 1.000)
            };
            \node[above] at ($(axis cs:0, 1.000)$) {\tiny 1.000};
            \node[above] at ($(axis cs:0.6, 1.000)$) {\tiny 1.000};
            
            \addplot [draw=light-gray, line width=0.7pt, fill=gray, error bars/.cd, y dir=both, y explicit, error bar style={draw=black}] coordinates {
                (0, 1.000)
                (0.6, 1.000)
            };
            \node[above] at ($(axis cs:0.16, 1.000)$) {\tiny 1.000};
            \node[above] at ($(axis cs:0.76, 1.000)$) {\tiny 1.000};
            \end{axis}
        \end{tikzpicture}
    \end{minipage}
    \hfill
    \begin{minipage}{0.235\textwidth}
        \centering
        \begin{tikzpicture}
            \centering
            \begin{axis}[
                height=4cm, width=5cm,
                /pgf/bar width=0.26cm,
                xmin=-0.2, xmax=0.8,
                axis x line*=bottom, axis y line*=left, enlarge x limits=true,
                xtick={0, 0.6},
                xticklabel style={yshift=-0.8mm, font=\scriptsize, align=center},
                ybar=3.8pt, clip=false,
                ymin=0, ymax=1, ytick={0, 0.25, 0.5, 0.75, 1.0}, yticklabels={0.00, 0.25, 0.50, 0.75, 1.00},
                ymajorgrids, major grid style={draw=black!20}, tick align=inside,
                yticklabel style={font=\footnotesize}, tickwidth=0pt,
                y axis line style={opacity=0},
                ylabel={\scriptsize Score (Functional Corr.)},
                y label style={at={(0.55, 1.2)}, rotate=-90},
                xticklabels={\shortstack[c]{Low-Detailed\\Prompt}, \shortstack[c]{High-Detailed\\Prompt}},
            ]
    
            \addplot [draw=light-gray, line width=0.7pt, fill=color-e, error bars/.cd, y dir=both, y explicit, error bar style={draw=black}] coordinates {
                (0, 0.733)
                (0.6, 0.854)
            };
            \node[above] at ($(axis cs:-0.15, 0.733)$) {\tiny 0.733};
            \node[above] at ($(axis cs:0.45, 0.854)$) {\tiny 0.854};
    
            \addplot [draw=light-gray, line width=0.7pt, fill=color-b, error bars/.cd, y dir=both, y explicit, error bar style={draw=black}] coordinates {
                (0, 0.884)
                (0.6, 0.979)
            };
            \node[above] at ($(axis cs:-0.01, 0.884)$) {\tiny 0.884};
            \node[above] at ($(axis cs:0.59, 0.979)$) {\tiny 0.979};
            
            \addplot [draw=light-gray, line width=0.7pt, fill=gray, error bars/.cd, y dir=both, y explicit, error bar style={draw=black}] coordinates {
                (0, 0.992)
                (0.6, 1.000)
            };
            \node[above] at ($(axis cs:0.15, 0.992)$) {\tiny 0.992};
            \node[above] at ($(axis cs:0.76, 1.000)$) {\tiny 1.000};
            \end{axis}
        \end{tikzpicture}
    \end{minipage}
    \caption{Comparison between base LLMs and \tool{} in generating syntactically and functionally correct code.}
    \label{fig:gensiac_func_correctness}
    \vspace{-3mm}
\end{figure}
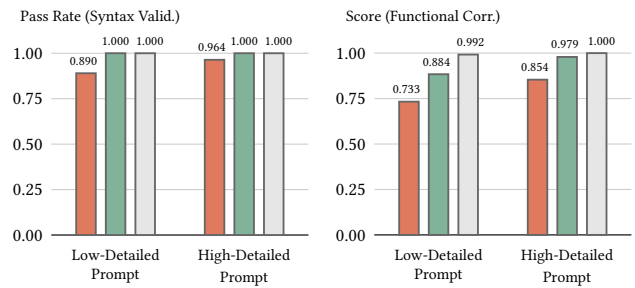

\paragraph{Syntactical and Functional Correctness}
In \cref{fig:gensiac_func_correctness}, we present a comparison of syntactical and functional correctness among the base \codellama{}, \gptfour{}, and \tool{} models.
As illustrated, both \tool{} and \gptfour{} achieve a perfect score of 1 for syntax validation in both low-detailed and high-detailed prompts, which is an improvement over the base \codellama{} (0.890 and 0.964).
In terms of functional correctness, \tool{} outperforms the base \codellama{} on both low-detailed and high-detailed prompts, with scores increasing from 0.733 to 0.884 and 0.854 to 0.979, respectively.
However, it is worth noting that the functional correctness for low-detailed prompts is still slightly lower for \tool{} compared to \gptfour{} by 0.108.
For high-detailed prompts, both \tool{} and \gptfour{} exhibit nearly perfect functional correctness.

\paragraph{Examples of IaC Security Weaknesses Detection}
We present one example generated by \tool{} (fine-tuning based on \codellama{}) during the evaluation mentioned above.
This example demonstrates how \tool{} incorporates code comments alongside security weaknesses in the code to accurately identify and highlight potential security issues.

\paragraph{Example I: CWE-798, Hard-Coded Secret}

\begin{lstlisting}[language=ruby, rulecolor=\color{black}]
- name: Create user
  openstack.cloud.identity_user:
     cloud: ""{{ cloud }}""
     state: present
     name: ansible_user
     (*@\textcolor{color-b}{\# Security smell! Hard-coded secret: please remove}@*)
     (*@\textcolor{color-b}{hard-coded secrets to prevent exposure...}@*)
     password: secret
\end{lstlisting}

\paragraph{Example II: CWE-284, Unrestricted IP Address}

\begin{lstlisting}[language=ruby, rulecolor=\color{black}]
(*@\textcolor{color-b}{\# Security smell! Unrestricted IP address: please do not}@*)
(*@\textcolor{color-b}{bind to 0.0.0.0...}@*)
memcached_listen_ip: ""0.0.0.0""
\end{lstlisting}

\paragraph{Example III: CWE-319, Use of HTTP Without TLS}

\begin{lstlisting}[language=ruby, rulecolor=\color{black}]
apt_repository ""apache-cassandra"" do
  (*@\textcolor{color-b}{\# Security smell! Use of HTTP without SSL/TLS: please use}@*)
  (*@\textcolor{color-b}{HTTPS instead of HTTP to prevent man-in-the-middle...}@*)
  uri ""http://www.apache.org/dist/cassandra/debian""
\end{lstlisting}

\paragraph{Example IV: CWE-494, No Integrity Check}

\begin{lstlisting}[language=ruby, rulecolor=\color{black}]
- name: create repo
  yum_repository:
    name: aap_installer
    description: aap_installer
    baseurl: ""file:///{{ aap_dir }}/bundle/el8/repos""
    (*@\textcolor{color-b}{\# Security smell! No integrity check: please verify...}@*)
    gpgcheck: false
\end{lstlisting}

\paragraph{Example V: CWE-478, Missing Default in Case Statement}

\begin{lstlisting}[language=ruby, rulecolor=\color{black}]
(*@\textcolor{color-b}{\# Security smell! Missing default in case statement: please}@*)
(*@\textcolor{color-b}{handle all input cases in conditionals...}@*)
case node['platform_family']
when 'debian'
  package 'libapache2-mod-wsgi'
when 'rhel', 'fedora', 'arch'
  package 'mod_wsgi' do
    notifies :run, 'execute[generate-module-list]', :immediately
  end
end
\end{lstlisting}

\subsection{Ablation Studies}
\label{sec:ablation_studies}

We present various ablation studies to validate the necessity of dataset creation for our \tool{}. 
The fine-tuning dataset for \tool{} consists of two parts: code generation data and code inspection data.
We conduct experiments by fine-tuning the base \codellama{}-13B with only code generation data and only code inspection data, comparing the performance with fine-tuning using the full dataset.
Throughout this section, we use consistent color notations: \basebox{} represents base \codellama{}, \partialbox{} represents fine-tuning with either code generation data or code inspection data, and \gensiacbox{} represents fine-tuning with the full \tool{} dataset.

\begin{figure}[htb]
    \begin{minipage}{0.235\textwidth}
        \centering
        \begin{tikzpicture}
            \centering
            \begin{axis}[
                height=4cm, width=5cm,
                /pgf/bar width=0.26cm,
                xmin=-0.2, xmax=0.8,
                axis x line*=bottom, axis y line*=left, enlarge x limits=true,
                xtick={0, 0.6},
                xticklabel style={yshift=-0.8mm, font=\scriptsize, align=center},
                ybar=3.8pt, clip=false,
                ymin=0, ymax=1, ytick={0, 0.25, 0.5, 0.75, 1.0}, yticklabels={0.00, 0.25, 0.50, 0.75, 1.00},
                ymajorgrids, major grid style={draw=black!20}, tick align=inside,
                yticklabel style={font=\footnotesize}, tickwidth=0pt,
                y axis line style={opacity=0},
                ylabel={\scriptsize F1 (Code Generation)},
                y label style={at={(0.55, 1.15)}, rotate=-90},
                xticklabels={\shortstack[c]{Train with\\code gen. data}, \shortstack[c]{Train with\\code insp. data}},
            ]
            
            \addplot [draw=light-gray, line width=0.7pt, fill=color-e, error bars/.cd, y dir=both, y explicit, error bar style={draw=black}] coordinates {
                (0, 0.276)
                (0.6, 0.276)
            };
            \node[above] at ($(axis cs:-0.155, 0.276)$) {\tiny 0.276};
            \node[above] at ($(axis cs:0.435, 0.276)$) {\tiny 0.276};
    
            \addplot [draw=light-gray, line width=0.7pt, fill=color-f, error bars/.cd, y dir=both, y explicit, error bar style={draw=black}] coordinates {
                (0, 0.667)
                (0.6, 0.244)
            };
            \node[above] at ($(axis cs:-0.01, 0.667)$) {\tiny 0.667};
            \node[above] at ($(axis cs:0.6, 0.244)$) {\tiny 0.244};
            
            \addplot [draw=light-gray, line width=0.7pt, fill=color-b, error bars/.cd, y dir=both, y explicit, error bar style={draw=black}] coordinates {
                (0, 0.771)
                (0.6, 0.771)
            };
            \node[above] at ($(axis cs:0.15, 0.771)$) {\tiny 0.771};
            \node[above] at ($(axis cs:0.75, 0.771)$) {\tiny 0.771};
            \end{axis}
        \end{tikzpicture}
    \end{minipage}
    \hfill
    \begin{minipage}{0.235\textwidth}
        \centering
        \begin{tikzpicture}
            \centering
            \begin{axis}[
                height=4cm, width=5cm,
                /pgf/bar width=0.26cm,
                xmin=-0.2, xmax=0.8,
                axis x line*=bottom, axis y line*=left, enlarge x limits=true,
                xtick={0, 0.6},
                xticklabel style={yshift=-0.8mm, font=\scriptsize, align=center},
                ybar=3.8pt, clip=false,
                ymin=0, ymax=1, ytick={0, 0.25, 0.5, 0.75, 1.0}, yticklabels={0.00, 0.25, 0.50, 0.75, 1.00},
                ymajorgrids, major grid style={draw=black!20}, tick align=inside,
                yticklabel style={font=\footnotesize}, tickwidth=0pt,
                y axis line style={opacity=0},
                ylabel={\scriptsize F1 (Code Inspection)},
                y label style={at={(0.55, 1.15)}, rotate=-90},
                xticklabels={\shortstack[c]{Train with\\code gen. data}, \shortstack[c]{Train with\\code insp. data}},
            ]
    
            \addplot [draw=light-gray, line width=0.7pt, fill=color-e, error bars/.cd, y dir=both, y explicit, error bar style={draw=black}] coordinates {
                (0, 0.303)
                (0.6, 0.303)
            };
            \node[above] at ($(axis cs:-0.15, 0.303)$) {\tiny 0.303};
            \node[above] at ($(axis cs:0.45, 0.303)$) {\tiny 0.303};
    
            \addplot [draw=light-gray, line width=0.7pt, fill=color-f, error bars/.cd, y dir=both, y explicit, error bar style={draw=black}] coordinates {
                (0, 0.138)
                (0.6, 0.850)
            };
            \node[above] at ($(axis cs:0, 0.138)$) {\tiny 0.138};
            \node[above] at ($(axis cs:0.58, 0.850)$) {\tiny 0.850};
            
            \addplot [draw=light-gray, line width=0.7pt, fill=color-b, error bars/.cd, y dir=both, y explicit, error bar style={draw=black}] coordinates {
                (0, 0.858)
                (0.6, 0.858)
            };
            \node[above] at ($(axis cs:0.15, 0.858)$) {\tiny 0.858};
            \node[above] at ($(axis cs:0.75, 0.858)$) {\tiny 0.858};
            \end{axis}
        \end{tikzpicture}
    \end{minipage}
    \caption{Overall F1-score for detecting nine security weaknesses using either code generation data or code inspection data for training.}
    \label{fig:gensiac_ablation}
    \vspace{-1mm}
\end{figure}
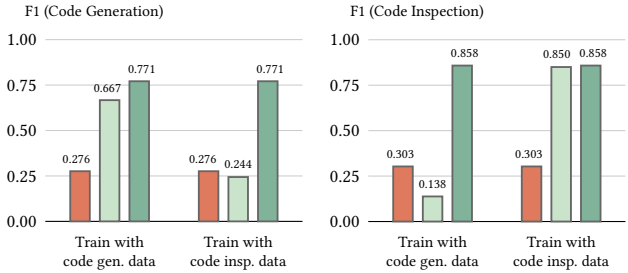

\paragraph{Necessity of Including Code Generation Data}
The results of fine-tuning \codellama{} with only code generation data are presented in \cref{fig:gensiac_ablation}.
As observed, when fine-tuning with only code generation data, the performance of security weakness identification during code generation improves the base model from 0.276 to 0.667, which is slightly lower than the model fine-tuned with the full \tool{} dataset (0.771). 
However, during code inspection, the performance is even worse than the base model (0.138 vs 0.303), which is significantly lower than the model fine-tuned with the full \tool{} dataset (0.858).

\paragraph{Necessity of Including Code Inspection Data}
The results of fine-tuning \codellama{} with only code inspection data are also shown in \cref{fig:gensiac_ablation}.
We observe a similar trend, where training only on code inspection data does not significantly improve the performance of security weakness detection during code generation (0.244 vs 0.276).
However, the performance during code inspection is on par with fine-tuning using the full dataset (0.850 vs 0.858). 
This shows that training only on code inspection data can improve the performance of security weakness detection during code inspection but does not enhance the performance during code generation.
Thus, it is necessary to include both code generation and inspection data to improve security weakness detection ability.

\subsection{\tool{} Generalizability Studies}
\label{sec:generalizability}

In this section, we evaluate the generalizability of \tool{}.
Throughout this section, we maintain consistent color notations: \basebox{} denotes base LLMs, \partialbox{} indicates fine-tuning with a portion of the \tool{} dataset, and \gensiacbox{} signifies fine-tuning with the complete \tool{} dataset.

\begin{figure}[htb]
    \begin{minipage}{0.49\textwidth}
        \centering
        \begin{tikzpicture}
            \centering
            \begin{axis}[
                height=4cm, width=9cm,
                /pgf/bar width=0.26cm,
                xmin=-0.2, xmax=2.2,
                axis x line*=bottom, axis y line*=left, enlarge x limits=true,
                xtick={0, 1, 2},
                xticklabel style={yshift=-0.8mm, font=\scriptsize, align=center},
                ybar=3.8pt, clip=false,
                ymin=0, ymax=1, ytick={0, 0.25, 0.5, 0.75, 1.0}, yticklabels={0.00, 0.25, 0.50, 0.75, 1.00},
                ymajorgrids, major grid style={draw=black!20}, tick align=inside,
                yticklabel style={font=\footnotesize}, tickwidth=0pt,
                y axis line style={opacity=0},
                ylabel={\scriptsize F1 (Code Generation)},
                y label style={at={(0.21, 1.15)}, rotate=-90},
                xticklabels={\shortstack[c]{\codellama{}\\Target: Ansible}, \shortstack[c]{\codellama{}\\Target: Chef}, \shortstack[c]{\codellama{}\\Target: Puppet}},
            ]
            
            \addplot [draw=light-gray, line width=0.7pt, fill=color-e, error bars/.cd, y dir=both, y explicit, error bar style={draw=black}] coordinates {
                (0, 0.371)
                (1, 0.121)
                (2, 0.200)
            };
            \node[above] at ($(axis cs:-0.17, 0.371)$) {\tiny 0.371};
            \node[above] at ($(axis cs:0.83, 0.121)$) {\tiny 0.121};
            \node[above] at ($(axis cs:1.83, 0.200)$) {\tiny 0.200};
    
            \addplot [draw=light-gray, line width=0.7pt, fill=color-f, error bars/.cd, y dir=both, y explicit, error bar style={draw=black}] coordinates {
                (0, 0.417)
                (1, 0.412)
                (2, 0.495)
            };
            \node[above] at ($(axis cs:0, 0.417)$) {\tiny 0.417};
            \node[above] at ($(axis cs:1, 0.412)$) {\tiny 0.412};
            \node[above] at ($(axis cs:2, 0.495)$) {\tiny 0.495};
            
            \addplot [draw=light-gray, line width=0.7pt, fill=color-b, error bars/.cd, y dir=both, y explicit, error bar style={draw=black}] coordinates {
                (0, 0.802)
                (1, 0.800)
                (2, 0.610)
            };
            \node[above] at ($(axis cs:0.15, 0.802)$) {\tiny 0.802};
            \node[above] at ($(axis cs:1.15, 0.800)$) {\tiny 0.800};
            \node[above] at ($(axis cs:2.15, 0.610)$) {\tiny 0.610};
            \end{axis}
        \end{tikzpicture}
    \end{minipage}
    \hfill
    \begin{minipage}{0.49\textwidth}
        \centering
        \begin{tikzpicture}
            \centering
            \begin{axis}[
                height=4cm, width=9cm,
                /pgf/bar width=0.26cm,
                xmin=-0.2, xmax=2.2,
                axis x line*=bottom, axis y line*=left, enlarge x limits=true,
                xtick={0, 1, 2},
                xticklabel style={yshift=-0.8mm, font=\scriptsize, align=center},
                ybar=3.8pt, clip=false,
                ymin=0, ymax=1, ytick={0, 0.25, 0.5, 0.75, 1.0}, yticklabels={0.00, 0.25, 0.50, 0.75, 1.00},
                ymajorgrids, major grid style={draw=black!20}, tick align=inside,
                yticklabel style={font=\footnotesize}, tickwidth=0pt,
                y axis line style={opacity=0},
                ylabel={\scriptsize F1 (Code Inspection)},
                y label style={at={(0.21, 1.15)}, rotate=-90},
                xticklabels={\shortstack[c]{\codellama{}\\Target: Ansible}, \shortstack[c]{\codellama{}\\Target: Chef}, \shortstack[c]{\codellama{}\\Target: Puppet}},
            ]
            
            \addplot [draw=light-gray, line width=0.7pt, fill=color-e, error bars/.cd, y dir=both, y explicit, error bar style={draw=black}] coordinates {
                (0, 0.387)
                (1, 0.116)
                (2, 0.339)
            };
            \node[above] at ($(axis cs:-0.18, 0.387)$) {\tiny 0.387};
            \node[above] at ($(axis cs:0.83, 0.116)$) {\tiny 0.116};
            \node[above] at ($(axis cs:1.83, 0.339)$) {\tiny 0.339};
            
            \addplot [draw=light-gray, line width=0.7pt, fill=color-f, error bars/.cd, y dir=both, y explicit, error bar style={draw=black}] coordinates {
                (0, 0.403)
                (1, 0.505)
                (2, 0.712)
            };
            \node[above] at ($(axis cs:0, 0.403)$) {\tiny 0.403};
            \node[above] at ($(axis cs:1, 0.505)$) {\tiny 0.505};
            \node[above] at ($(axis cs:2, 0.712)$) {\tiny 0.712};
    
            \addplot [draw=light-gray, line width=0.7pt, fill=color-b, error bars/.cd, y dir=both, y explicit, error bar style={draw=black}] coordinates {
                (0, 0.874)
                (1, 0.872)
                (2, 0.805)
            };
            \node[above] at ($(axis cs:0.15, 0.874)$) {\tiny 0.874};
            \node[above] at ($(axis cs:1.15, 0.872)$) {\tiny 0.872};
            \node[above] at ($(axis cs:2.15, 0.805)$) {\tiny 0.805};
            \end{axis}
        \end{tikzpicture}
    \end{minipage}
    \caption{Cross-language validation for security smell detection during code generation and inspection.}
    \label{fig:gensiac_cross_language}
    \vspace{-1mm}
\end{figure}
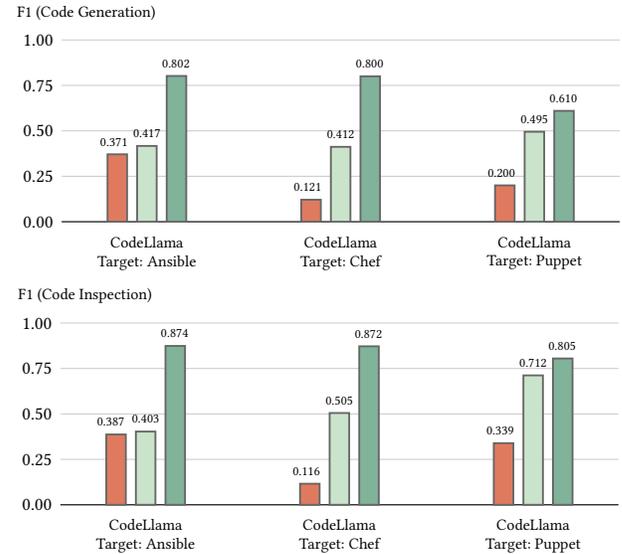

\paragraph{Cross-Language Performance}
Considering the wide variety of IaC languages available, it is always challenging to simultaneously support all of them for recognizing security weaknesses. 
Researchers continuously work to improve IaC language support, as demonstrated by Saavedra \textit{et al.}, who expanded their tool from three IaC languages \cite{DBLP:conf/kbse/Saavedra022} to six \cite{saavedra2023polyglot}.
Utilizing LLMs, such as \codellama{} \cite{DBLP:journals/corr/abs-2308-12950}, pretrained on diverse datasets containing multiple languages, suggests that \tool{} might be capable of recognizing security weaknesses across different IaC languages.
To evaluate this, we perform cross-language validation experiments with \tool{}.
Our dataset includes three IaC languages: Ansible, Chef, and Puppet.
We specifically fine-tune \codellama{}-13B on two IaC languages and assess its performance on the third.
For example, we fine-tune on Ansible and Chef, and then test on Puppet.
We repeat this process three times for each task (\cref{fig:gensiac_cross_language}).

The results indicate that \tool{} can generalize to another IaC language, while the performance for all languages improves compared to the base model.
However, the performance for all languages is still inferior to training on the entire \tool{} dataset during both code generation and inspection.
Specifically, during both code generation and inspection processes, the performance of the model significantly improves for two IaC languages, Chef and Puppet, compared to base models, while for Ansible, the improvement is relatively minor.
In particular, when targeting Ansible, the base model only improves from 0.371 and 0.387 to 0.417 and 0.403 for code generation and inspection, respectively.
When targeting Puppet, the performance significantly improves, becoming comparable to the performance of training on the complete \tool{} dataset, with 0.495 and 0.712 compared to 0.610 and 0.805.
This highlights \tool{}'s ability to generalize knowledge across languages while emphasizing the significance of targeted language training for optimal performance.
It is worth noting that although the \tool{} effectively generalizes, targeted language training can still be employed to further optimize performance for specific languages.

\section{Discussion}

In this section, we address the limitations and potential future work. 
Firstly, \tool{} does not currently support all IaC security weaknesses, such as \emph{CWE-276: Insecure File Permissions}, as discussed in \cref{sec:base_llm}.
This limitation stems from the fact that existing tools only support specific IaC security weaknesses, and we utilize a state-of-the-art tool to generate the instruction fine-tuning dataset for \tool{}. 
However, our proposed approach for fine-tuning base LLMs with the \tool{} dataset is designed to be easily extendable.
In the future, as more IaC security weaknesses are incorporated, researchers can conveniently expand our dataset with new samples of security weaknesses. 
Practitioners can then use the extended instruction fine-tuning dataset to fine-tune LLMs, enabling the detection of new IaC security issues.
Secondly, we observe that base LLMs exhibit limited performance in IaC security weakness detection during code generation and inspection. 
Fine-tuning with our \tool{} dataset using LoRA significantly enhances the performance of LLMs. 
However, it remains unclear whether other parameter-efficient fine-tuning techniques could achieve superior performance in IaC security detection. 
Consequently, we recommend exploring alternative parameter-efficient fine-tuning methods, such as adapter tuning, prefix tuning, and prompt tuning approaches, to fine-tune and evaluate the performance of the models.

\section{Conclusion}

In conclusion, this paper addresses two significant problems related to the security of IaC code generated by LLMs and their ability to recognize IaC security weaknesses.
To tackle these issues, we first conduct a comprehensive evaluation of base LLMs in recognizing major IaC security weaknesses during the generation and inspection of IaC code.
We then propose a novel instruction fine-tuning dataset, \tool{}, designed to enhance LLMs' IaC code generation and IaC security weakness detection capabilities. 
The results demonstrate that fine-tuned LLMs achieve strong performance in IaC security weakness detection and functional correctness, significantly surpassing the performance of base LLMs without fine-tuning.

\bibliography{bibliography}


\begin{thebibliography}{46}


\ifx \showCODEN    \undefined \def \showCODEN     #1{\unskip}     \fi
\ifx \showDOI      \undefined \def \showDOI       #1{#1}\fi
\ifx \showISBNx    \undefined \def \showISBNx     #1{\unskip}     \fi
\ifx \showISBNxiii \undefined \def \showISBNxiii  #1{\unskip}     \fi
\ifx \showISSN     \undefined \def \showISSN      #1{\unskip}     \fi
\ifx \showLCCN     \undefined \def \showLCCN      #1{\unskip}     \fi
\ifx \shownote     \undefined \def \shownote      #1{#1}          \fi
\ifx \showarticletitle \undefined \def \showarticletitle #1{#1}   \fi
\ifx \showURL      \undefined \def \showURL       {\relax}        \fi
\providecommand\bibfield[2]{#2}
\providecommand\bibinfo[2]{#2}
\providecommand\natexlab[1]{#1}
\providecommand\showeprint[2][]{arXiv:#2}

\bibitem[gen(2024a)]%
        {genIaC}
 \bibinfo{year}{2024}\natexlab{a}.
\newblock \bibinfo{title}{AI-Generated Infrastructure-as-Code: The Good, the Bad and the Ugly}.
\newblock
\newblock
\urldef\tempurl%
\url{https://doi.org/blog/ai-generated-infrastructure-as-code-the-good-the-bad-and-the-ugly}
\showDOI{\tempurl}
\newblock
\shownote{Accessed: 2024-01-10}.


\bibitem[AIa(2024)]%
        {AIaC}
 \bibinfo{year}{2024}\natexlab{}.
\newblock \bibinfo{title}{AIaC}.
\newblock \bibinfo{howpublished}{\url{https://github.com/gofireflyio/aiac}}.
\newblock
\newblock
\shownote{Accessed: 2024-01-10}.


\bibitem[cod(2024)]%
        {codewhisperer}
 \bibinfo{year}{2024}\natexlab{}.
\newblock \bibinfo{title}{Amazon CodeWhisperer}.
\newblock \bibinfo{howpublished}{\url{https://aws.amazon.com/codewhisperer/}}.
\newblock
\newblock
\shownote{Accessed: 2024-01-10}.


\bibitem[dns(2024)]%
        {dns_outage}
 \bibinfo{year}{2024}\natexlab{}.
\newblock \bibinfo{title}{DNS Outage Post Mortem}.
\newblock \bibinfo{howpublished}{\url{https://github.blog/2014-01-18-dns-outage-post-mortem/}}.
\newblock
\newblock
\shownote{Accessed: 2024-01-10}.


\bibitem[cop(2024)]%
        {copilot}
 \bibinfo{year}{2024}\natexlab{}.
\newblock \bibinfo{title}{GitHub Copilot}.
\newblock \bibinfo{howpublished}{\url{https://github.com/features/copilot}}.
\newblock
\newblock
\shownote{Accessed: 2024-01-10}.


\bibitem[gen(2024b)]%
        {genIaC2}
 \bibinfo{year}{2024}\natexlab{b}.
\newblock \bibinfo{title}{Mastering the Future: Evaluating LLM-Generated Data Architectures leveraging IaC technologies}.
\newblock \bibinfo{howpublished}{Towards Data Science}.
\newblock
\urldef\tempurl%
\url{https://doi.org/mastering-the-future-evaluating-llm-generated-data-architectures-leveraging-iac-technologies-dee75302a355}
\showDOI{\tempurl}
\newblock
\shownote{Accessed: 2024-01-10}.


\bibitem[gho(2024)]%
        {ghostwriter}
 \bibinfo{year}{2024}\natexlab{}.
\newblock \bibinfo{title}{Replit GhostWriter}.
\newblock \bibinfo{howpublished}{\url{https://replit.com/ai}}.
\newblock
\newblock
\shownote{Accessed: 2024-01-10}.


\bibitem[Bhandari et~al\mbox{.}(2021)]%
        {bhandari2021cvefixes}
\bibfield{author}{\bibinfo{person}{Guru Bhandari}, \bibinfo{person}{Amara Naseer}, {and} \bibinfo{person}{Leon Moonen}.} \bibinfo{year}{2021}\natexlab{}.
\newblock \showarticletitle{CVEfixes: automated collection of vulnerabilities and their fixes from open-source software}. In \bibinfo{booktitle}{\emph{Proceedings of the 17th International Conference on Predictive Models and Data Analytics in Software Engineering}}. \bibinfo{pages}{30--39}.
\newblock


\bibitem[Brown et~al\mbox{.}(2020)]%
        {brown2020language}
\bibfield{author}{\bibinfo{person}{Tom Brown}, \bibinfo{person}{Benjamin Mann}, \bibinfo{person}{Nick Ryder}, \bibinfo{person}{Melanie Subbiah}, \bibinfo{person}{Jared~D Kaplan}, \bibinfo{person}{Prafulla Dhariwal}, \bibinfo{person}{Arvind Neelakantan}, \bibinfo{person}{Pranav Shyam}, \bibinfo{person}{Girish Sastry}, \bibinfo{person}{Amanda Askell}, {et~al\mbox{.}}} \bibinfo{year}{2020}\natexlab{}.
\newblock \showarticletitle{Language models are few-shot learners}.
\newblock \bibinfo{journal}{\emph{Advances in neural information processing systems}}  \bibinfo{volume}{33} (\bibinfo{year}{2020}), \bibinfo{pages}{1877--1901}.
\newblock


\bibitem[Chen et~al\mbox{.}(2021)]%
        {DBLP:journals/corr/abs-2107-03374}
\bibfield{author}{\bibinfo{person}{Mark Chen}, \bibinfo{person}{Jerry Tworek}, \bibinfo{person}{Heewoo Jun}, \bibinfo{person}{Qiming Yuan}, \bibinfo{person}{Henrique~Pond{\'{e}} de Oliveira~Pinto}, \bibinfo{person}{Jared Kaplan}, \bibinfo{person}{Harrison Edwards}, \bibinfo{person}{Yuri Burda}, \bibinfo{person}{Nicholas Joseph}, \bibinfo{person}{Greg Brockman}, \bibinfo{person}{Alex Ray}, \bibinfo{person}{Raul Puri}, \bibinfo{person}{Gretchen Krueger}, \bibinfo{person}{Michael Petrov}, \bibinfo{person}{Heidy Khlaaf}, \bibinfo{person}{Girish Sastry}, \bibinfo{person}{Pamela Mishkin}, \bibinfo{person}{Brooke Chan}, \bibinfo{person}{Scott Gray}, \bibinfo{person}{Nick Ryder}, \bibinfo{person}{Mikhail Pavlov}, \bibinfo{person}{Alethea Power}, \bibinfo{person}{Lukasz Kaiser}, \bibinfo{person}{Mohammad Bavarian}, \bibinfo{person}{Clemens Winter}, \bibinfo{person}{Philippe Tillet}, \bibinfo{person}{Felipe~Petroski Such}, \bibinfo{person}{Dave Cummings}, \bibinfo{person}{Matthias Plappert}, \bibinfo{person}{Fotios
  Chantzis}, \bibinfo{person}{Elizabeth Barnes}, \bibinfo{person}{Ariel Herbert{-}Voss}, \bibinfo{person}{William~Hebgen Guss}, \bibinfo{person}{Alex Nichol}, \bibinfo{person}{Alex Paino}, \bibinfo{person}{Nikolas Tezak}, \bibinfo{person}{Jie Tang}, \bibinfo{person}{Igor Babuschkin}, \bibinfo{person}{Suchir Balaji}, \bibinfo{person}{Shantanu Jain}, \bibinfo{person}{William Saunders}, \bibinfo{person}{Christopher Hesse}, \bibinfo{person}{Andrew~N. Carr}, \bibinfo{person}{Jan Leike}, \bibinfo{person}{Joshua Achiam}, \bibinfo{person}{Vedant Misra}, \bibinfo{person}{Evan Morikawa}, \bibinfo{person}{Alec Radford}, \bibinfo{person}{Matthew Knight}, \bibinfo{person}{Miles Brundage}, \bibinfo{person}{Mira Murati}, \bibinfo{person}{Katie Mayer}, \bibinfo{person}{Peter Welinder}, \bibinfo{person}{Bob McGrew}, \bibinfo{person}{Dario Amodei}, \bibinfo{person}{Sam McCandlish}, \bibinfo{person}{Ilya Sutskever}, {and} \bibinfo{person}{Wojciech Zaremba}.} \bibinfo{year}{2021}\natexlab{}.
\newblock \showarticletitle{Evaluating Large Language Models Trained on Code}.
\newblock \bibinfo{journal}{\emph{CoRR}}  \bibinfo{volume}{abs/2107.03374} (\bibinfo{year}{2021}).
\newblock
\showeprint[arXiv]{2107.03374}
\urldef\tempurl%
\url{https://arxiv.org/abs/2107.03374}
\showURL{%
\tempurl}


\bibitem[Chen et~al\mbox{.}(2023)]%
        {chen2023diversevul}
\bibfield{author}{\bibinfo{person}{Yizheng Chen}, \bibinfo{person}{Zhoujie Ding}, \bibinfo{person}{Lamya Alowain}, \bibinfo{person}{Xinyun Chen}, {and} \bibinfo{person}{David Wagner}.} \bibinfo{year}{2023}\natexlab{}.
\newblock \showarticletitle{Diversevul: A new vulnerable source code dataset for deep learning based vulnerability detection}. In \bibinfo{booktitle}{\emph{Proceedings of the 26th International Symposium on Research in Attacks, Intrusions and Defenses}}. \bibinfo{pages}{654--668}.
\newblock


\bibitem[Chiang et~al\mbox{.}(2023)]%
        {vicuna2023}
\bibfield{author}{\bibinfo{person}{Wei-Lin Chiang}, \bibinfo{person}{Zhuohan Li}, \bibinfo{person}{Zi Lin}, \bibinfo{person}{Ying Sheng}, \bibinfo{person}{Zhanghao Wu}, \bibinfo{person}{Hao Zhang}, \bibinfo{person}{Lianmin Zheng}, \bibinfo{person}{Siyuan Zhuang}, \bibinfo{person}{Yonghao Zhuang}, \bibinfo{person}{Joseph~E. Gonzalez}, \bibinfo{person}{Ion Stoica}, {and} \bibinfo{person}{Eric~P. Xing}.} \bibinfo{year}{2023}\natexlab{}.
\newblock \bibinfo{title}{Vicuna: An Open-Source Chatbot Impressing GPT-4 with 90\%* ChatGPT Quality}.
\newblock
\newblock
\urldef\tempurl%
\url{https://lmsys.org/blog/2023-03-30-vicuna/}
\showURL{%
\tempurl}


\bibitem[Ding et~al\mbox{.}(2024)]%
        {ding2024vulnerability}
\bibfield{author}{\bibinfo{person}{Yangruibo Ding}, \bibinfo{person}{Yanjun Fu}, \bibinfo{person}{Omniyyah Ibrahim}, \bibinfo{person}{Chawin Sitawarin}, \bibinfo{person}{Xinyun Chen}, \bibinfo{person}{Basel Alomair}, \bibinfo{person}{David Wagner}, \bibinfo{person}{Baishakhi Ray}, {and} \bibinfo{person}{Yizheng Chen}.} \bibinfo{year}{2024}\natexlab{}.
\newblock \showarticletitle{Vulnerability Detection with Code Language Models: How Far Are We?}. In \bibinfo{booktitle}{\emph{2025 IEEE/ACM 47th International Conference on Software Engineering (ICSE)}}. IEEE Computer Society, \bibinfo{pages}{469--481}.
\newblock


\bibitem[Fan et~al\mbox{.}(2020)]%
        {fan2020ac}
\bibfield{author}{\bibinfo{person}{Jiahao Fan}, \bibinfo{person}{Yi Li}, \bibinfo{person}{Shaohua Wang}, {and} \bibinfo{person}{Tien~N Nguyen}.} \bibinfo{year}{2020}\natexlab{}.
\newblock \showarticletitle{A C/C++ code vulnerability dataset with code changes and CVE summaries}. In \bibinfo{booktitle}{\emph{Proceedings of the 17th International Conference on Mining Software Repositories}}. \bibinfo{pages}{508--512}.
\newblock


\bibitem[Fried et~al\mbox{.}(2022)]%
        {DBLP:journals/corr/abs-2204-05999}
\bibfield{author}{\bibinfo{person}{Daniel Fried}, \bibinfo{person}{Armen Aghajanyan}, \bibinfo{person}{Jessy Lin}, \bibinfo{person}{Sida Wang}, \bibinfo{person}{Eric Wallace}, \bibinfo{person}{Freda Shi}, \bibinfo{person}{Ruiqi Zhong}, \bibinfo{person}{Wen{-}tau Yih}, \bibinfo{person}{Luke Zettlemoyer}, {and} \bibinfo{person}{Mike Lewis}.} \bibinfo{year}{2022}\natexlab{}.
\newblock \showarticletitle{InCoder: {A} Generative Model for Code Infilling and Synthesis}.
\newblock \bibinfo{journal}{\emph{CoRR}}  \bibinfo{volume}{abs/2204.05999} (\bibinfo{year}{2022}).
\newblock
\urldef\tempurl%
\url{https://doi.org/10.48550/ARXIV.2204.05999}
\showDOI{\tempurl}
\showeprint[arXiv]{2204.05999}


\bibitem[He and Vechev(2023)]%
        {DBLP:conf/ccs/HeV23}
\bibfield{author}{\bibinfo{person}{Jingxuan He} {and} \bibinfo{person}{Martin~T. Vechev}.} \bibinfo{year}{2023}\natexlab{}.
\newblock \showarticletitle{Large Language Models for Code: Security Hardening and Adversarial Testing}. In \bibinfo{booktitle}{\emph{Proceedings of the 2023 {ACM} {SIGSAC} Conference on Computer and Communications Security, {CCS} 2023, Copenhagen, Denmark, November 26-30, 2023}}, \bibfield{editor}{\bibinfo{person}{Weizhi Meng}, \bibinfo{person}{Christian~Damsgaard Jensen}, \bibinfo{person}{Cas Cremers}, {and} \bibinfo{person}{Engin Kirda}} (Eds.). \bibinfo{publisher}{{ACM}}, \bibinfo{pages}{1865--1879}.
\newblock
\urldef\tempurl%
\url{https://doi.org/10.1145/3576915.3623175}
\showDOI{\tempurl}


\bibitem[Houlsby et~al\mbox{.}(2019)]%
        {houlsby2019parameter}
\bibfield{author}{\bibinfo{person}{Neil Houlsby}, \bibinfo{person}{Andrei Giurgiu}, \bibinfo{person}{Stanislaw Jastrzebski}, \bibinfo{person}{Bruna Morrone}, \bibinfo{person}{Quentin De~Laroussilhe}, \bibinfo{person}{Andrea Gesmundo}, \bibinfo{person}{Mona Attariyan}, {and} \bibinfo{person}{Sylvain Gelly}.} \bibinfo{year}{2019}\natexlab{}.
\newblock \showarticletitle{Parameter-efficient transfer learning for NLP}. In \bibinfo{booktitle}{\emph{International Conference on Machine Learning}}. PMLR, \bibinfo{pages}{2790--2799}.
\newblock


\bibitem[Hu et~al\mbox{.}(2022)]%
        {DBLP:conf/iclr/HuSWALWWC22}
\bibfield{author}{\bibinfo{person}{Edward~J. Hu}, \bibinfo{person}{Yelong Shen}, \bibinfo{person}{Phillip Wallis}, \bibinfo{person}{Zeyuan Allen{-}Zhu}, \bibinfo{person}{Yuanzhi Li}, \bibinfo{person}{Shean Wang}, \bibinfo{person}{Lu Wang}, {and} \bibinfo{person}{Weizhu Chen}.} \bibinfo{year}{2022}\natexlab{}.
\newblock \showarticletitle{LoRA: Low-Rank Adaptation of Large Language Models}. In \bibinfo{booktitle}{\emph{The Tenth International Conference on Learning Representations, {ICLR} 2022, Virtual Event, April 25-29, 2022}}. \bibinfo{publisher}{OpenReview.net}.
\newblock
\urldef\tempurl%
\url{https://openreview.net/forum?id=nZeVKeeFYf9}
\showURL{%
\tempurl}


\bibitem[Khoury et~al\mbox{.}(2023)]%
        {DBLP:journals/corr/abs-2304-09655}
\bibfield{author}{\bibinfo{person}{Rapha{\"{e}}l Khoury}, \bibinfo{person}{Anderson~R. Avila}, \bibinfo{person}{Jacob Brunelle}, {and} \bibinfo{person}{Baba~Mamadou Camara}.} \bibinfo{year}{2023}\natexlab{}.
\newblock \showarticletitle{How Secure is Code Generated by ChatGPT?}
\newblock \bibinfo{journal}{\emph{CoRR}}  \bibinfo{volume}{abs/2304.09655} (\bibinfo{year}{2023}).
\newblock
\urldef\tempurl%
\url{https://doi.org/10.48550/ARXIV.2304.09655}
\showDOI{\tempurl}
\showeprint[arXiv]{2304.09655}


\bibitem[Lester et~al\mbox{.}(2021)]%
        {lester2021power}
\bibfield{author}{\bibinfo{person}{Brian Lester}, \bibinfo{person}{Rami Al-Rfou}, {and} \bibinfo{person}{Noah Constant}.} \bibinfo{year}{2021}\natexlab{}.
\newblock \showarticletitle{The power of scale for parameter-efficient prompt tuning}.
\newblock \bibinfo{journal}{\emph{arXiv preprint arXiv:2104.08691}} (\bibinfo{year}{2021}).
\newblock


\bibitem[Li et~al\mbox{.}(2023)]%
        {li2023starcoder}
\bibfield{author}{\bibinfo{person}{Raymond Li}, \bibinfo{person}{Loubna~Ben Allal}, \bibinfo{person}{Yangtian Zi}, \bibinfo{person}{Niklas Muennighoff}, \bibinfo{person}{Denis Kocetkov}, \bibinfo{person}{Chenghao Mou}, \bibinfo{person}{Marc Marone}, \bibinfo{person}{Christopher Akiki}, \bibinfo{person}{Jia Li}, \bibinfo{person}{Jenny Chim}, {et~al\mbox{.}}} \bibinfo{year}{2023}\natexlab{}.
\newblock \showarticletitle{StarCoder: may the source be with you!}
\newblock \bibinfo{journal}{\emph{arXiv preprint arXiv:2305.06161}} (\bibinfo{year}{2023}).
\newblock


\bibitem[Li and Liang(2021)]%
        {li2021prefix}
\bibfield{author}{\bibinfo{person}{Xiang~Lisa Li} {and} \bibinfo{person}{Percy Liang}.} \bibinfo{year}{2021}\natexlab{}.
\newblock \showarticletitle{Prefix-tuning: Optimizing continuous prompts for generation}.
\newblock \bibinfo{journal}{\emph{arXiv preprint arXiv:2101.00190}} (\bibinfo{year}{2021}).
\newblock


\bibitem[Li et~al\mbox{.}(2025)]%
        {li2025out}
\bibfield{author}{\bibinfo{person}{Yikun Li}, \bibinfo{person}{Ngoc~Tan Bui}, \bibinfo{person}{Ting Zhang}, \bibinfo{person}{Martin Weyssow}, \bibinfo{person}{Chengran Yang}, \bibinfo{person}{Xin Zhou}, \bibinfo{person}{Jinfeng Jiang}, \bibinfo{person}{Junkai Chen}, \bibinfo{person}{Huihui Huang}, \bibinfo{person}{Huu~Hung Nguyen}, {et~al\mbox{.}}} \bibinfo{year}{2025}\natexlab{}.
\newblock \showarticletitle{Out of Distribution, Out of Luck: How Well Can LLMs Trained on Vulnerability Datasets Detect Top 25 CWE Weaknesses?}
\newblock \bibinfo{journal}{\emph{arXiv preprint arXiv:2507.21817}} (\bibinfo{year}{2025}).
\newblock


\bibitem[Li et~al\mbox{.}(2024)]%
        {li2024cleanvul}
\bibfield{author}{\bibinfo{person}{Yikun Li}, \bibinfo{person}{Ting Zhang}, \bibinfo{person}{Ratnadira Widyasari}, \bibinfo{person}{Yan~Naing Tun}, \bibinfo{person}{Huu~Hung Nguyen}, \bibinfo{person}{Tan Bui}, \bibinfo{person}{Ivana~Clairine Irsan}, \bibinfo{person}{Yiran Cheng}, \bibinfo{person}{Xiang Lan}, \bibinfo{person}{Han~Wei Ang}, {et~al\mbox{.}}} \bibinfo{year}{2024}\natexlab{}.
\newblock \showarticletitle{CleanVul: Automatic Function-Level Vulnerability Detection in Code Commits Using LLM Heuristics}.
\newblock \bibinfo{journal}{\emph{arXiv preprint arXiv:2411.17274}} (\bibinfo{year}{2024}).
\newblock


\bibitem[Luo et~al\mbox{.}(2023)]%
        {luo2023wizardcoder}
\bibfield{author}{\bibinfo{person}{Ziyang Luo}, \bibinfo{person}{Can Xu}, \bibinfo{person}{Pu Zhao}, \bibinfo{person}{Qingfeng Sun}, \bibinfo{person}{Xiubo Geng}, \bibinfo{person}{Wenxiang Hu}, \bibinfo{person}{Chongyang Tao}, \bibinfo{person}{Jing Ma}, \bibinfo{person}{Qingwei Lin}, {and} \bibinfo{person}{Daxin Jiang}.} \bibinfo{year}{2023}\natexlab{}.
\newblock \showarticletitle{WizardCoder: Empowering Code Large Language Models with Evol-Instruct}.
\newblock \bibinfo{journal}{\emph{arXiv preprint arXiv:2306.08568}} (\bibinfo{year}{2023}).
\newblock


\bibitem[Morris(2016)]%
        {morris2016infrastructure}
\bibfield{author}{\bibinfo{person}{Kief Morris}.} \bibinfo{year}{2016}\natexlab{}.
\newblock \bibinfo{booktitle}{\emph{Infrastructure as code: managing servers in the cloud}}.
\newblock \bibinfo{publisher}{" O'Reilly Media, Inc."}.
\newblock


\bibitem[Nijkamp et~al\mbox{.}(2023)]%
        {nijkamp2023codegen2}
\bibfield{author}{\bibinfo{person}{Erik Nijkamp}, \bibinfo{person}{Hiroaki Hayashi}, \bibinfo{person}{Caiming Xiong}, \bibinfo{person}{Silvio Savarese}, {and} \bibinfo{person}{Yingbo Zhou}.} \bibinfo{year}{2023}\natexlab{}.
\newblock \showarticletitle{Codegen2: Lessons for training llms on programming and natural languages}.
\newblock \bibinfo{journal}{\emph{arXiv preprint arXiv:2305.02309}} (\bibinfo{year}{2023}).
\newblock


\bibitem[Opdebeeck et~al\mbox{.}(2023)]%
        {opdebeeck2023control}
\bibfield{author}{\bibinfo{person}{Ruben Opdebeeck}, \bibinfo{person}{Ahmed Zerouali}, {and} \bibinfo{person}{Coen~De Roover}.} \bibinfo{year}{2023}\natexlab{}.
\newblock \showarticletitle{Control and Data Flow in Security Smell Detection for Infrastructure as Code: Is It Worth the Effort?}. In \bibinfo{booktitle}{\emph{20th {IEEE/ACM} International Conference on Mining Software Repositories, {MSR} 2023, Melbourne, Australia, May 15-16, 2023}}. \bibinfo{publisher}{{IEEE}}, \bibinfo{pages}{534--545}.
\newblock
\urldef\tempurl%
\url{https://doi.org/10.1109/MSR59073.2023.00079}
\showDOI{\tempurl}


\bibitem[OpenAI(2023)]%
        {DBLP:journals/corr/abs-2303-08774}
\bibfield{author}{\bibinfo{person}{OpenAI}.} \bibinfo{year}{2023}\natexlab{}.
\newblock \showarticletitle{{GPT-4} Technical Report}.
\newblock \bibinfo{journal}{\emph{CoRR}}  \bibinfo{volume}{abs/2303.08774} (\bibinfo{year}{2023}).
\newblock
\urldef\tempurl%
\url{https://doi.org/10.48550/ARXIV.2303.08774}
\showDOI{\tempurl}
\showeprint[arXiv]{2303.08774}


\bibitem[Pearce et~al\mbox{.}(2022)]%
        {pearce2022asleep}
\bibfield{author}{\bibinfo{person}{Hammond Pearce}, \bibinfo{person}{Baleegh Ahmad}, \bibinfo{person}{Benjamin Tan}, \bibinfo{person}{Brendan Dolan{-}Gavitt}, {and} \bibinfo{person}{Ramesh Karri}.} \bibinfo{year}{2022}\natexlab{}.
\newblock \showarticletitle{Asleep at the Keyboard? Assessing the Security of GitHub Copilot's Code Contributions}. In \bibinfo{booktitle}{\emph{43rd {IEEE} Symposium on Security and Privacy, {SP} 2022, San Francisco, CA, USA, May 22-26, 2022}}. \bibinfo{publisher}{{IEEE}}, \bibinfo{pages}{754--768}.
\newblock
\urldef\tempurl%
\url{https://doi.org/10.1109/SP46214.2022.9833571}
\showDOI{\tempurl}


\bibitem[Rahman et~al\mbox{.}(2019)]%
        {rahman2019seven}
\bibfield{author}{\bibinfo{person}{Akond Rahman}, \bibinfo{person}{Chris Parnin}, {and} \bibinfo{person}{Laurie~A. Williams}.} \bibinfo{year}{2019}\natexlab{}.
\newblock \showarticletitle{The seven sins: security smells in infrastructure as code scripts}. In \bibinfo{booktitle}{\emph{Proceedings of the 41st International Conference on Software Engineering, {ICSE} 2019, Montreal, QC, Canada, May 25-31, 2019}}, \bibfield{editor}{\bibinfo{person}{Joanne~M. Atlee}, \bibinfo{person}{Tevfik Bultan}, {and} \bibinfo{person}{Jon Whittle}} (Eds.). \bibinfo{publisher}{{IEEE} / {ACM}}, \bibinfo{pages}{164--175}.
\newblock
\urldef\tempurl%
\url{https://doi.org/10.1109/ICSE.2019.00033}
\showDOI{\tempurl}


\bibitem[Rahman et~al\mbox{.}(2021)]%
        {DBLP:journals/tosem/RahmanRPW21}
\bibfield{author}{\bibinfo{person}{Akond Rahman}, \bibinfo{person}{Md.~Rayhanur Rahman}, \bibinfo{person}{Chris Parnin}, {and} \bibinfo{person}{Laurie~A. Williams}.} \bibinfo{year}{2021}\natexlab{}.
\newblock \showarticletitle{Security Smells in Ansible and Chef Scripts: {A} Replication Study}.
\newblock \bibinfo{journal}{\emph{{ACM} Trans. Softw. Eng. Methodol.}} \bibinfo{volume}{30}, \bibinfo{number}{1} (\bibinfo{year}{2021}), \bibinfo{pages}{3:1--3:31}.
\newblock
\urldef\tempurl%
\url{https://doi.org/10.1145/3408897}
\showDOI{\tempurl}


\bibitem[Rahman and Williams(2021)]%
        {DBLP:journals/ieeesp/RahmanW21}
\bibfield{author}{\bibinfo{person}{Akond Rahman} {and} \bibinfo{person}{Laurie~A. Williams}.} \bibinfo{year}{2021}\natexlab{}.
\newblock \showarticletitle{Different Kind of Smells: Security Smells in Infrastructure as Code Scripts}.
\newblock \bibinfo{journal}{\emph{{IEEE} Secur. Priv.}} \bibinfo{volume}{19}, \bibinfo{number}{3} (\bibinfo{year}{2021}), \bibinfo{pages}{33--41}.
\newblock
\urldef\tempurl%
\url{https://doi.org/10.1109/MSEC.2021.3065190}
\showDOI{\tempurl}


\bibitem[Rozi{\`{e}}re et~al\mbox{.}(2023)]%
        {DBLP:journals/corr/abs-2308-12950}
\bibfield{author}{\bibinfo{person}{Baptiste Rozi{\`{e}}re}, \bibinfo{person}{Jonas Gehring}, \bibinfo{person}{Fabian Gloeckle}, \bibinfo{person}{Sten Sootla}, \bibinfo{person}{Itai Gat}, \bibinfo{person}{Xiaoqing~Ellen Tan}, \bibinfo{person}{Yossi Adi}, \bibinfo{person}{Jingyu Liu}, \bibinfo{person}{Tal Remez}, \bibinfo{person}{J{\'{e}}r{\'{e}}my Rapin}, \bibinfo{person}{Artyom Kozhevnikov}, \bibinfo{person}{Ivan Evtimov}, \bibinfo{person}{Joanna Bitton}, \bibinfo{person}{Manish Bhatt}, \bibinfo{person}{Cristian Canton{-}Ferrer}, \bibinfo{person}{Aaron Grattafiori}, \bibinfo{person}{Wenhan Xiong}, \bibinfo{person}{Alexandre D{\'{e}}fossez}, \bibinfo{person}{Jade Copet}, \bibinfo{person}{Faisal Azhar}, \bibinfo{person}{Hugo Touvron}, \bibinfo{person}{Louis Martin}, \bibinfo{person}{Nicolas Usunier}, \bibinfo{person}{Thomas Scialom}, {and} \bibinfo{person}{Gabriel Synnaeve}.} \bibinfo{year}{2023}\natexlab{}.
\newblock \showarticletitle{Code Llama: Open Foundation Models for Code}.
\newblock \bibinfo{journal}{\emph{CoRR}}  \bibinfo{volume}{abs/2308.12950} (\bibinfo{year}{2023}).
\newblock
\urldef\tempurl%
\url{https://doi.org/10.48550/ARXIV.2308.12950}
\showDOI{\tempurl}
\showeprint[arXiv]{2308.12950}


\bibitem[Saavedra and Ferreira(2022)]%
        {DBLP:conf/kbse/Saavedra022}
\bibfield{author}{\bibinfo{person}{Nuno Saavedra} {and} \bibinfo{person}{Jo{\~{a}}o~F. Ferreira}.} \bibinfo{year}{2022}\natexlab{}.
\newblock \showarticletitle{{GLITCH:} Automated Polyglot Security Smell Detection in Infrastructure as Code}. In \bibinfo{booktitle}{\emph{37th {IEEE/ACM} International Conference on Automated Software Engineering, {ASE} 2022, Rochester, MI, USA, October 10-14, 2022}}. \bibinfo{publisher}{{ACM}}, \bibinfo{pages}{47:1--47:12}.
\newblock
\urldef\tempurl%
\url{https://doi.org/10.1145/3551349.3556945}
\showDOI{\tempurl}


\bibitem[Saavedra et~al\mbox{.}(2023)]%
        {saavedra2023polyglot}
\bibfield{author}{\bibinfo{person}{Nuno Saavedra}, \bibinfo{person}{Jo{\~{a}}o Gon{\c{c}}alves}, \bibinfo{person}{Miguel Henriques}, \bibinfo{person}{Jo{\~{a}}o~F. Ferreira}, {and} \bibinfo{person}{Alexandra Mendes}.} \bibinfo{year}{2023}\natexlab{}.
\newblock \showarticletitle{Polyglot Code Smell Detection for Infrastructure as Code with {GLITCH}}. In \bibinfo{booktitle}{\emph{38th {IEEE/ACM} International Conference on Automated Software Engineering, {ASE} 2023, Luxembourg, September 11-15, 2023}}. \bibinfo{publisher}{{IEEE}}, \bibinfo{pages}{2042--2045}.
\newblock
\urldef\tempurl%
\url{https://doi.org/10.1109/ASE56229.2023.00162}
\showDOI{\tempurl}


\bibitem[Sandoval et~al\mbox{.}(2023)]%
        {DBLP:conf/uss/SandovalPNKGD23}
\bibfield{author}{\bibinfo{person}{Gustavo Sandoval}, \bibinfo{person}{Hammond Pearce}, \bibinfo{person}{Teo Nys}, \bibinfo{person}{Ramesh Karri}, \bibinfo{person}{Siddharth Garg}, {and} \bibinfo{person}{Brendan Dolan{-}Gavitt}.} \bibinfo{year}{2023}\natexlab{}.
\newblock \showarticletitle{Lost at {C:} {A} User Study on the Security Implications of Large Language Model Code Assistants}. In \bibinfo{booktitle}{\emph{32nd {USENIX} Security Symposium, {USENIX} Security 2023, Anaheim, CA, USA, August 9-11, 2023}}, \bibfield{editor}{\bibinfo{person}{Joseph~A. Calandrino} {and} \bibinfo{person}{Carmela Troncoso}} (Eds.). \bibinfo{publisher}{{USENIX} Association}, \bibinfo{pages}{2205--2222}.
\newblock
\urldef\tempurl%
\url{https://www.usenix.org/conference/usenixsecurity23/presentation/sandoval}
\showURL{%
\tempurl}


\bibitem[Touvron et~al\mbox{.}(2023)]%
        {touvron2023llama}
\bibfield{author}{\bibinfo{person}{Hugo Touvron}, \bibinfo{person}{Louis Martin}, \bibinfo{person}{Kevin Stone}, \bibinfo{person}{Peter Albert}, \bibinfo{person}{Amjad Almahairi}, \bibinfo{person}{Yasmine Babaei}, \bibinfo{person}{Nikolay Bashlykov}, \bibinfo{person}{Soumya Batra}, \bibinfo{person}{Prajjwal Bhargava}, \bibinfo{person}{Shruti Bhosale}, {et~al\mbox{.}}} \bibinfo{year}{2023}\natexlab{}.
\newblock \showarticletitle{Llama 2: Open foundation and fine-tuned chat models}.
\newblock \bibinfo{journal}{\emph{arXiv preprint arXiv:2307.09288}} (\bibinfo{year}{2023}).
\newblock


\bibitem[Vaithilingam et~al\mbox{.}(2022)]%
        {DBLP:conf/chi/Vaithilingam0G22}
\bibfield{author}{\bibinfo{person}{Priyan Vaithilingam}, \bibinfo{person}{Tianyi Zhang}, {and} \bibinfo{person}{Elena~L. Glassman}.} \bibinfo{year}{2022}\natexlab{}.
\newblock \showarticletitle{Expectation vs. Experience: Evaluating the Usability of Code Generation Tools Powered by Large Language Models}. In \bibinfo{booktitle}{\emph{{CHI} '22: {CHI} Conference on Human Factors in Computing Systems, New Orleans, LA, USA, 29 April 2022 - 5 May 2022, Extended Abstracts}}, \bibfield{editor}{\bibinfo{person}{Simone D.~J. Barbosa}, \bibinfo{person}{Cliff Lampe}, \bibinfo{person}{Caroline Appert}, {and} \bibinfo{person}{David~A. Shamma}} (Eds.). \bibinfo{publisher}{{ACM}}, \bibinfo{pages}{332:1--332:7}.
\newblock
\urldef\tempurl%
\url{https://doi.org/10.1145/3491101.3519665}
\showDOI{\tempurl}


\bibitem[Vaswani et~al\mbox{.}(2017)]%
        {DBLP:conf/nips/VaswaniSPUJGKP17}
\bibfield{author}{\bibinfo{person}{Ashish Vaswani}, \bibinfo{person}{Noam Shazeer}, \bibinfo{person}{Niki Parmar}, \bibinfo{person}{Jakob Uszkoreit}, \bibinfo{person}{Llion Jones}, \bibinfo{person}{Aidan~N. Gomez}, \bibinfo{person}{Lukasz Kaiser}, {and} \bibinfo{person}{Illia Polosukhin}.} \bibinfo{year}{2017}\natexlab{}.
\newblock \showarticletitle{Attention is All you Need}. In \bibinfo{booktitle}{\emph{Advances in Neural Information Processing Systems 30: Annual Conference on Neural Information Processing Systems 2017, December 4-9, 2017, Long Beach, CA, {USA}}}, \bibfield{editor}{\bibinfo{person}{Isabelle Guyon}, \bibinfo{person}{Ulrike von Luxburg}, \bibinfo{person}{Samy Bengio}, \bibinfo{person}{Hanna~M. Wallach}, \bibinfo{person}{Rob Fergus}, \bibinfo{person}{S.~V.~N. Vishwanathan}, {and} \bibinfo{person}{Roman Garnett}} (Eds.). \bibinfo{pages}{5998--6008}.
\newblock


\bibitem[Weyssow et~al\mbox{.}(2025)]%
        {weyssow2025r2vul}
\bibfield{author}{\bibinfo{person}{Martin Weyssow}, \bibinfo{person}{Chengran Yang}, \bibinfo{person}{Junkai Chen}, \bibinfo{person}{Ratnadira Widyasari}, \bibinfo{person}{Ting Zhang}, \bibinfo{person}{Huihui Huang}, \bibinfo{person}{Huu~Hung Nguyen}, \bibinfo{person}{Yan~Naing Tun}, \bibinfo{person}{Tan Bui}, \bibinfo{person}{Yikun Li}, {et~al\mbox{.}}} \bibinfo{year}{2025}\natexlab{}.
\newblock \showarticletitle{R2Vul: Learning to Reason about Software Vulnerabilities with Reinforcement Learning and Structured Reasoning Distillation}.
\newblock \bibinfo{journal}{\emph{arXiv preprint arXiv:2504.04699}} (\bibinfo{year}{2025}).
\newblock


\bibitem[Xu et~al\mbox{.}(2022)]%
        {DBLP:journals/corr/abs-2202-13169}
\bibfield{author}{\bibinfo{person}{Frank~F. Xu}, \bibinfo{person}{Uri Alon}, \bibinfo{person}{Graham Neubig}, {and} \bibinfo{person}{Vincent~J. Hellendoorn}.} \bibinfo{year}{2022}\natexlab{}.
\newblock \showarticletitle{A Systematic Evaluation of Large Language Models of Code}.
\newblock \bibinfo{journal}{\emph{CoRR}}  \bibinfo{volume}{abs/2202.13169} (\bibinfo{year}{2022}).
\newblock
\showeprint[arXiv]{2202.13169}
\urldef\tempurl%
\url{https://arxiv.org/abs/2202.13169}
\showURL{%
\tempurl}


\bibitem[Yetistiren et~al\mbox{.}(2022)]%
        {DBLP:conf/promise/YetistirenOT22}
\bibfield{author}{\bibinfo{person}{Burak Yetistiren}, \bibinfo{person}{Isik Ozsoy}, {and} \bibinfo{person}{Eray Tuzun}.} \bibinfo{year}{2022}\natexlab{}.
\newblock \showarticletitle{Assessing the quality of GitHub copilot's code generation}. In \bibinfo{booktitle}{\emph{Proceedings of the 18th International Conference on Predictive Models and Data Analytics in Software Engineering, {PROMISE} 2022, Singapore, Singapore, 17 November 2022}}, \bibfield{editor}{\bibinfo{person}{Shane McIntosh}, \bibinfo{person}{Weiyi Shang}, {and} \bibinfo{person}{Gema Rodr{\'{\i}}guez{-}P{\'{e}}rez}} (Eds.). \bibinfo{publisher}{{ACM}}, \bibinfo{pages}{62--71}.
\newblock
\urldef\tempurl%
\url{https://doi.org/10.1145/3558489.3559072}
\showDOI{\tempurl}


\bibitem[Yuan et~al\mbox{.}(2023)]%
        {yuan2023evaluating}
\bibfield{author}{\bibinfo{person}{Zhiqiang Yuan}, \bibinfo{person}{Junwei Liu}, \bibinfo{person}{Qiancheng Zi}, \bibinfo{person}{Mingwei Liu}, \bibinfo{person}{Xin Peng}, {and} \bibinfo{person}{Yiling Lou}.} \bibinfo{year}{2023}\natexlab{}.
\newblock \showarticletitle{Evaluating instruction-tuned large language models on code comprehension and generation}.
\newblock \bibinfo{journal}{\emph{arXiv preprint arXiv:2308.01240}} (\bibinfo{year}{2023}).
\newblock


\bibitem[Zhao et~al\mbox{.}(2023)]%
        {DBLP:journals/corr/abs-2303-18223}
\bibfield{author}{\bibinfo{person}{Wayne~Xin Zhao}, \bibinfo{person}{Kun Zhou}, \bibinfo{person}{Junyi Li}, \bibinfo{person}{Tianyi Tang}, \bibinfo{person}{Xiaolei Wang}, \bibinfo{person}{Yupeng Hou}, \bibinfo{person}{Yingqian Min}, \bibinfo{person}{Beichen Zhang}, \bibinfo{person}{Junjie Zhang}, \bibinfo{person}{Zican Dong}, \bibinfo{person}{Yifan Du}, \bibinfo{person}{Chen Yang}, \bibinfo{person}{Yushuo Chen}, \bibinfo{person}{Zhipeng Chen}, \bibinfo{person}{Jinhao Jiang}, \bibinfo{person}{Ruiyang Ren}, \bibinfo{person}{Yifan Li}, \bibinfo{person}{Xinyu Tang}, \bibinfo{person}{Zikang Liu}, \bibinfo{person}{Peiyu Liu}, \bibinfo{person}{Jian{-}Yun Nie}, {and} \bibinfo{person}{Ji{-}Rong Wen}.} \bibinfo{year}{2023}\natexlab{}.
\newblock \showarticletitle{A Survey of Large Language Models}.
\newblock \bibinfo{journal}{\emph{CoRR}}  \bibinfo{volume}{abs/2303.18223} (\bibinfo{year}{2023}).
\newblock
\urldef\tempurl%
\url{https://doi.org/10.48550/ARXIV.2303.18223}
\showDOI{\tempurl}
\showeprint[arXiv]{2303.18223}


\bibitem[Zhuo(2023)]%
        {DBLP:journals/corr/abs-2304-14317}
\bibfield{author}{\bibinfo{person}{Terry~Yue Zhuo}.} \bibinfo{year}{2023}\natexlab{}.
\newblock \showarticletitle{Large Language Models Are State-of-the-Art Evaluators of Code Generation}.
\newblock \bibinfo{journal}{\emph{CoRR}}  \bibinfo{volume}{abs/2304.14317} (\bibinfo{year}{2023}).
\newblock
\urldef\tempurl%
\url{https://doi.org/10.48550/ARXIV.2304.14317}
\showDOI{\tempurl}
\showeprint[arXiv]{2304.14317}


\end{thebibliography}
\bibliographystyle{ACM-Reference-Format}

\end{document}